\documentclass[twocolumn]{aastex7}

\usepackage[colorinlistoftodos]{todonotes}
\usepackage{amsmath}
\usepackage{pbox}

\usepackage{url}
\makeatletter
\g@addto@macro{\UrlBreaks}{\UrlOrds}
\makeatother

\usepackage{enumitem}
\setlist{leftmargin=*}

\begin{document}
\journalinfo{Accepted for publication in the Astronomical Journal, \today}
\shortauthors{White \& Peek}
\shorttitle{Hubble Image Similarity Project}

\title{The Hubble Image Similarity Project}

\correspondingauthor{Richard L. White}
\email{rlw@stsci.edu}

\author[0000-0002-9194-2807]{Richard L. White}
\email{rlw@stsci.edu}
\affiliation{Space Telescope Science Institute,
3700 San Martin Drive,
Baltimore MD 21218, USA}

\author[0000-0003-4797-7030]{J. E. G. Peek}
\email{jegpeek@stsci.edu}
\affiliation{Space Telescope Science Institute,
3700 San Martin Drive,
Baltimore MD 21218, USA}
\affiliation{Department of Physics \& Astronomy,
Johns Hopkins University,
3400 N. Charles Street,
Baltimore, MD 21218, USA}

\begin{abstract}

We have created a large database of similarity information between
sub-regions of \emph{Hubble} Space Telescope images.  These data can
be used to assess the accuracy of image search algorithms based
on computer vision methods.  The images were compared by humans in
a citizen science project, where they were asked to select similar
images from a comparison sample.  We utilized the Amazon Mechanical
Turk system to pay our reviewers a fair wage for their work.
Nearly 850,000 comparison measurements have been analyzed to construct
a similarity distance matrix between all the pairs of images.  We
describe the algorithm used to extract a robust distance matrix
from the (sometimes noisy) user reviews.  The results are very
impressive: the data capture similarity between images based on
morphology, texture, and other details that are sometimes difficult
even to describe in words (e.g., dusty absorption bands with sharp
edges).  The collective visual wisdom of our citizen scientists
matches the accuracy of the trained eye, with even subtle differences
among images faithfully reflected in the distances.

\end{abstract}



\section{Introduction} \label{sec:intro}

Archives of astronomical images allow users to find images by metadata: camera,
filter, exposure time, sky position, observation epoch, etc.  Searches based on
the scientific value of the image are sometimes possible when the scientific
motivation for the project (via principal investigator, proposal abstract,
target category) or the scientific results (via links to the text of papers that
utilized the data) are made possible.  Catalogs of the objects contained in
those images \citep[e.g., the \emph{Hubble} Source Catalog,][]{whitmore2016}
provide a limited search of the data itself: position, magnitude, diameter,
position angle, S\'ersic index.

\begin{figure*}
	\centering
    \includegraphics[width=\textwidth]{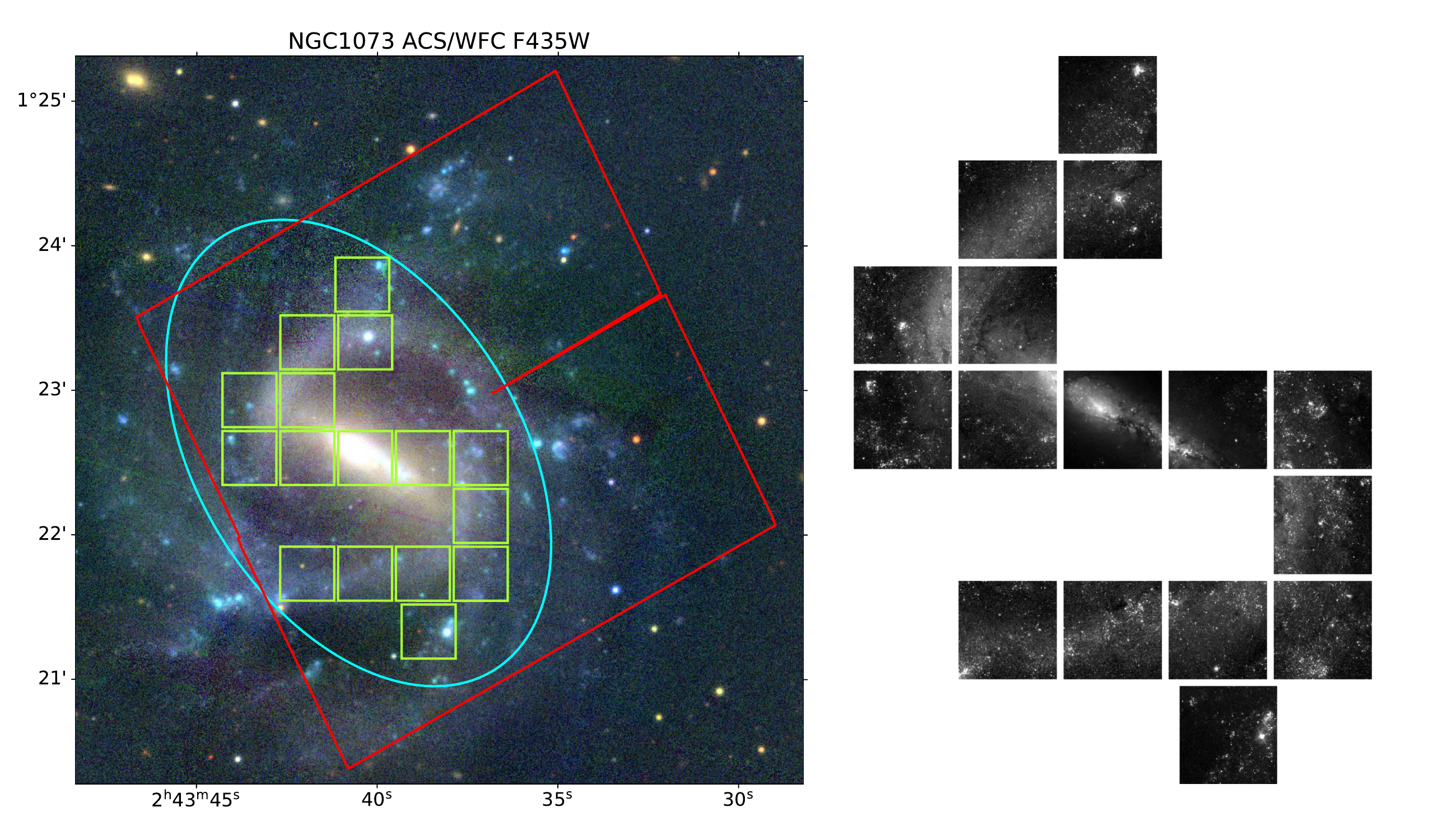}
	\caption{Example showing cutout selections within the
	elliptical region defined for NGC~1073.  The left panel
	shows the NGC catalog ellipse (cyan), the footprint for
	the HLA observation (red), and the regions for the selected
	cutout images (green).  The background image is from the
	Pan-STARRS1 \emph{gri} filters.  The right panel shows the HST cutout
	images.
	The removal of cutouts having low entropy leaves only
	images in regions with relatively high contrast.  For this object
	the images trace the galaxy's spiral arms.
\label{fig:ngc1073}}
\end{figure*}

None of these search techniques provides much assistance in finding images that
show more complex structures.  Spatial morphology has long played a key role in
the discovery and classification of astronomical objects, from the era of
artistic astronomy \citep[e.g.][]{Rosse1850}, to studies of the shapes of galaxies in the high-redshift
universe \citep[e.g.][]{hstfaint}, to computer-vision and machine-learning powered morphology tools for
diffuse media \citep[][]{pb19}. Model fitting provides an efficient approach
for extracting quantitative morphological characteristics for relatively compact
objects.  The vast number of papers using the SDSS and 2MASS data attests to
the value of such catalogs \citep{sdss,2mass}.
But model fitting is less useful for extended objects having complex,
highly variable structures, especially in the common case where there is no
obvious choice for a good model.  What is the correct model for the shape of substructures in dusty star-forming galaxies (e.g., Fig.~\ref{fig:ngc1073})?

Given the currently limited search capabilities we look to answer this question:
given a complex image, how can we find all the images in the archive that look
like it?  New algorithms from machine learning and artificial intelligence provide powerful tools to address this ``search-by-image'' capability.

The human eye and brain can discern both global and local patterns in complex
images, providing a discovery tool that remains unmatched in its overall
flexibility and power. The eye can identify important patterns and can use them
to group objects and structures into physically meaningful clusters. However,
the quantitative use of qualitative judgments from visual examination can be
challenging.

Machine Vision approaches, especially Convolutional Neural Networks (CNNs) and
variations of them, have opened the door to fast, quantitative image
classification and clustering that captures arbitrary features in images
\citep[for a review see][]{lecun2015}.
These neural network algorithms can be used
as the core of tools that provide repeatable results amenable to statistical
analysis, and that can be shared like other data analysis tools so they can be
improved and compared.  In particular, advances in transfer and self-supervised
learning \citep{ssl} within the deep learning paradigm have made it possible to develop robust search-by-image systems that do not depend on heavily labeled data as are
typically required for the first generation of convolutional networks.

Indeed, prototype systems have been developed both for astrophysics and earth
science for searching databases not by image meta-data, but using images
themselves.  \cite{peek2020} used transfer learning combined with dimensionality
reduction techniques to apply convolutional neural networks trained on
terrestrial images to find structures such as edge-on galaxies with dust lanes,
star fields, and rich scenes of star formation.  \cite{hayat2021} applied
self-supervised learning to images of galaxies to extract morphological
information.  \cite{seo2023} used a convolutional autoencoder on galaxy images
to find similar images. \cite{astroml} applied self-supervised learning to satellite images of
the earth to similar ends. 

A key problem with these systems is that they are hard to validate. While some
labeled images do exist for each scientific subdomain, there are few if any
datasets available that have been specifically designed to have image similarity
measures.  How do we determine which of the available approaches work best, and
how can we quantitatively assess their performance?  At its core the question is
whether the groups of images identified by one algorithm are more similar than
the groups from another algorithm.

\begin{deluxetable*}{cp{0.6\textwidth}}
\tablecolumns{2}
\tablecaption{HISP Lexicon\label{tab:lexicon}}
\tablehead{
        \colhead{Terminology} &
		\colhead{Description}
}
\tablewidth{0pt}
\startdata
	\emph{cutout image} & $224\times224$ pixel grayscale JPEG image (binned down from $448\times448$
	original Hubble pixels), sometimes simply called an \emph{image} \\
	\emph{parent sample} & Collection of $\sim20,000$ cutout images from \emph{Hubble}
	observations of NGC objects \\
	\emph{selected sample} & 2,098 image subset from the parent sample utilized for reviews \\
	\emph{page} & Reference image plus a group of comparison images, the unit of reviews \\
	\emph{collection} & Complete set of all review pages defining one phase of the review \\
	\emph{golden question} & Page where the reference image is also included among the comparison images,
	used for assessment of reviewer error rate and biases \\
	\emph{reviewer} & Citizen scientist who participated in the project \\
\enddata
\end{deluxetable*}

This work, \emph{The Hubble Image Similarity Project (HISP)}, aims to create a large database of similarity information between segments of \emph{Hubble} Space Telescope (HST) images.  The images were compared by humans in a citizen science project, where
they were asked to select similar images from a comparison sample.

We could have utilized a review system such as Zooniverse \citep{lintott2008}
that relies on volunteer efforts by citizen scientists.  Many projects in
astronomy have had great success using this approach
\citep[e.g.,][]{zevin2017,christiansen2018,mahabal2019}.  However, we questioned
the ethics of relying on unpaid labor during the global economic crisis that was
spawned by the Covid-19 pandemic.  Instead, we utilized the Amazon Mechanical
Turk (AMT) system to pay our reviewers a fair wage for their work.  AMT has
infrastructure in place to support micropayments of a few cents per task and
allows tasks to be restricted to a particular group of reviewers.  Creating a
similar system from scratch for this project would have presented major
bureaucratic and funding obstacles, but working via AMT allowed us to use
existing funding paths already in use for Amazon web services.

To maximize the local economic impact of our project, we chose members of the
community near the Space Telescope Science Institute in Baltimore who were
impacted by the economic downturn.  Using a defined and known set of reviewers
has many advantages for our project.  We had orientation sessions at the
beginning of the project to explain the data and the goals, and we had regular
sessions along the way to keep in touch with our team.  Participants could
contact us directly with questions about the data; we also encouraged them to
ask questions about the scientific content of the images and our project goals,
which increased their engagement with the project.  Our reviewers became highly
experienced in the visual aspects and mechanics of doing our image similarity
comparisons, which improved their efficiency and accuracy in the assigned tasks.

The comparison measurements have been analyzed to compute a distance matrix
between all the pairs of images, and that distance matrix can subsequently be
utilized to assess the accuracy of algorithms based on computer vision methods.
The image similarity matrix shows that the collective visual wisdom of our
neighbors matches the accuracy of the trained eye, with even subtle differences
among images faithfully reflected in the distances.

This paper describes the creation of the image sample, the design of the
reviewer tests, and the resulting datasets.  Readers interested mainly in an
overview of the results can go directly to section~\ref{sec:analysis}.
Section~\ref{sec:sample} describes the creation of the HISP image sample.
Section~\ref{sec:comparisons} lays out the three-phase procedure used for the
citizen science comparisons of the images, including the use of ``golden
questions'' with known answers that are included to assess reviewer accuracy.
Section~\ref{sec:results} summarizes the results of those reviews.
Section~\ref{sec:analysis} analyzes the resulting distance matrix and presents
evidence supporting the high quality of the results. The details of our approach
to computing a similarity distance matrix from the review data are described in
Appendix~\ref{sec:similarity-matrix}.  The final Section~\ref{sec:summary}
summarizes our results and discusses our plans for future work.

All the data products created for this paper are available in
the Mikulski Archive for Space Telescopes (MAST) as High Level Science Products:
\dataset[doi:~10.17909/0q3g-by85]{\doi{10.17909/0q3g-by85}}.
Table~\ref{tab:lexicon} summarizes the terminology used in the paper.

\section{The HISP Image Sample} \label{sec:sample}

We created the sample of images for our project by cross-matching the OpenNGC
catalog\footnote{\url{https://github.com/mattiaverga/OpenNGC}} \citep{openngc} with
images from the \emph{Hubble} Legacy Archive (HLA).  We chose fields
with NGC objects to get a wide variety of morphological structures.
The OpenNGC catalog was selected because it includes elliptical
sizes for the objects.  We selected objects with major axis sizes
between 1 and 4 arcmin; such objects are well-resolved by HST while
being reasonably well-matched to the $\sim3$~arcmin fields of view
of the HST cameras.  We included all objects regardless of the object
classification, so they include small and large galaxies, planetary nebulae, star clusters,
star formation regions, and more.  There are 6,527 NGC objects in that size range.

We matched the NGC object list to the HLA observations using the
\texttt{MastCaomCrossmatch}
service\footnote{\url{https://mast.stsci.edu/api/v0/_services.html}},
which selects all the HLA observations that fall within a circular
region around a list of target positions.  The list of observations was
further restricted by applying some additional constraints to improve the quality
and consistency of the images:

\begin{itemize}
	\item Exclude grisms, prisms, ramp filters, quadrant filters, polarization observations,
		multi-filter color images, and ultraviolet (UV) filters.

	\item Exclude multi-visit mosaic images, which are much less uniform than single-visit images.

	\item Keep only observations from the major HST cameras:
		the Advanced Camera for Surveys Wide Field and High
		Resolution Cameras (ACS/WFC, ACS/HRC), the Wide-Field
		Camera 3 UV/visible and infrared cameras (WFC3/UVIS,
		WFC3/IR) and the Wide-Field Planetary Camera 2
		(WFPC2).

	\item Reject CCD images having only a single exposure to avoid cosmic ray contamination.

	\item Require a minimum exposure time of 200 seconds to eliminate some low signal-to-noise short exposure images.
\end{itemize}

The initial search used a 4~arcmin diameter circle around each
object. The spatial search was refined by computing the overlap
between the NGC object's elliptical footprint and the HLA image
footprints (Fig.~\ref{fig:ngc1073}).  We used the astropy \texttt{regions}
module\footnote{\url{https://astropy-regions.readthedocs.io/}} to
facilitate the overlap computations \citep{astropy}.
The complete list of \emph{Hubble} images in our sample can be found in MAST:
\dataset[doi:~10.17909/3gmn-c234]{\doi{10.17909/3gmn-c234}}.

For each of the images that remain after this filtering process, we generate one
or more cutout images.  The cutouts all have a fixed size of $448\times448$
pixels in the HLA combined images.  We bin the cutouts to $224\times224$ pixels
for display to our reviewers.  Note that the pixel scale for the cutouts is
different depending on the camera, since the pixel size ranges from 0.1~arcsec
for WFPC2 to 0.025~arcsec for ACS/HRC.  The portion of the NGC object covered by
each cutout varies widely as well, with some smaller objects having only a
single cutout image that shows essentially the entire object, while large
objects are sliced up into numerous pieces that show only a portion of the
object (e.g., a spiral arm or a portion of a nebular cloud).

Note that the image cutout regions are not centered on object features
such as the galaxy nucleus or spiral arms.  In some cases they may be centered
by chance, particular for smaller objects or observations that have only a
single cutout.  But the positions of the cutout images are determined entirely
by the mechanical process of overlaying the NGC object's ellipse with the HST
image boundaries, as illustrated in Figure~\ref{fig:ngc1073}.

We use the standard HLA contrast mapping algorithm to generate an 8-bit
grayscale JPEG image for each cutout.  The algorithm sorts the FITS image pixel
values and trims the brightest and faintest 0.25\% of the pixels to get the
display range.  An arc hyperbolic sine transform is applied to improve the
contrast for fainter structures \citep{lupton2004}.  The result is a standard
set of 8-bit grayscale images suitable for viewing on web pages.

We use grayscale images rather than multi-color images constructed from
different filters because the HST sky coverage in different filters is highly
inhomogeneous.  There are regions of the sky with more than a dozen filters
available, and there are regions with only a single filter.  There are more than
80 instrument/filter combinations within our sample, and the variety of color
images is even greater.  We concluded that the best approach was to use only
single-filter grayscale images so that the sample retains a semblance of
uniformity.

In the end, our \emph{parent sample} includes 19,916 cutout images extracted
from 4,465 observations covering 1,071 NGC objects.  We refer to this as the
parent sample because it is much larger than we could accommodate in our HISP
reviews, where we wanted to do thorough comparisons between all possible pairs
of images (see section~\ref{sec:comparisons} for details).  We reduced the
number of cutouts in our \emph{selected sample} by a factor of $\sim10$ by
applying some additional restrictions:

\begin{figure*}
	\centering
    \includegraphics[width=\textwidth]{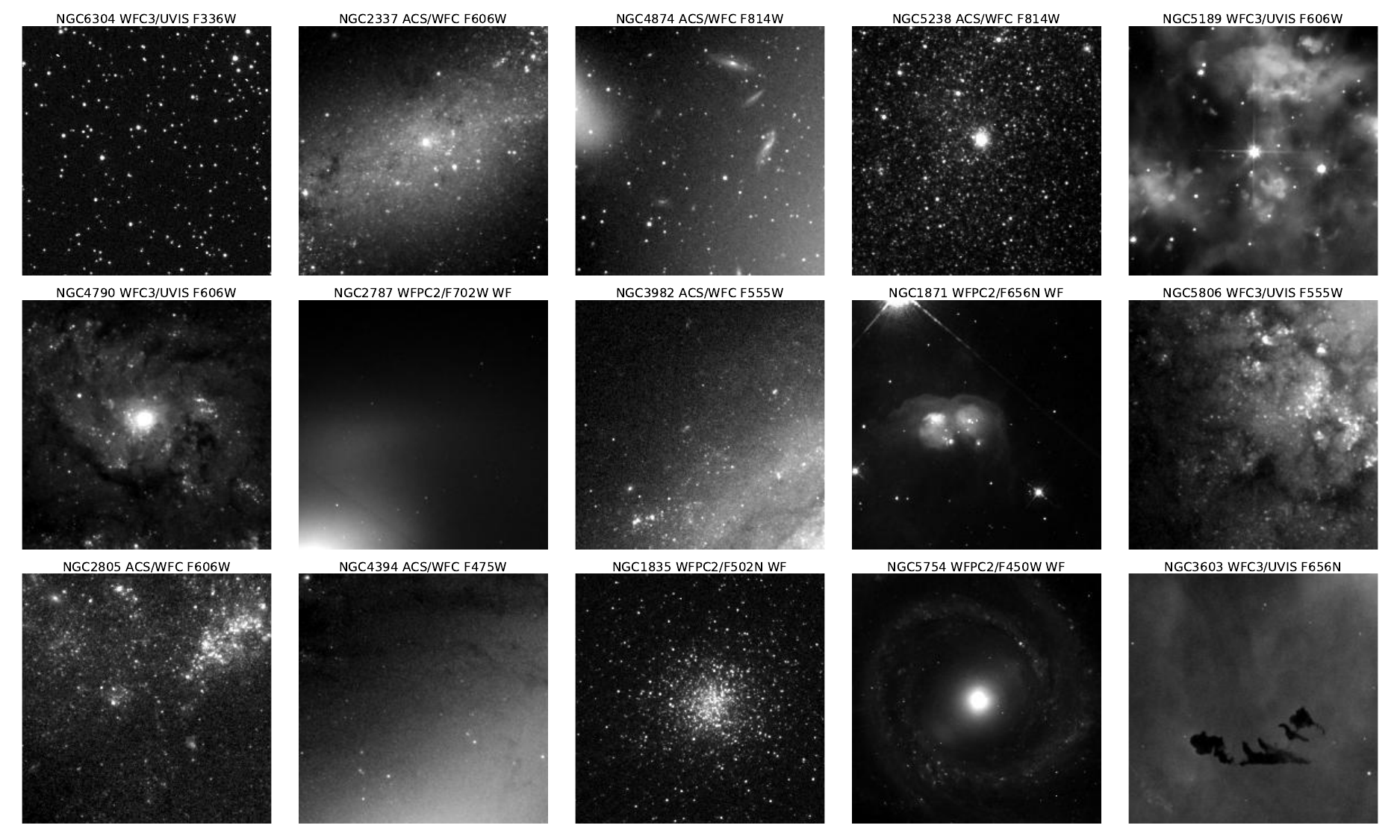}
	\caption{Sample of HST images used in HISP comparison tests.
	The titles indicate the object name, camera, and filter used
	for the observation.  Each image is a $448\times448$ pixel
	region from the HST (HLA) image.  There is a great variety
	in the morphology and texture of the images, which was the
	goal of the image selection process.  The entire image set
	includes a total of 2,098 images similar to these.
\label{fig:samples}}
\end{figure*}

\begin{itemize}
	\item We computed the global and local Shannon entropy\footnote{
		Using functions \texttt{skimage.measure.shannon\_entropy}
		and \texttt{skimage.filters.rank.entropy} from
		the Python \texttt{scikit-image} module.}
		for each cutout image.  These measure the amount of global and
		local contrast in the images, with values that range from 0 (for a constant image)
		to 8 (for a completely random image) for our 8-bit images.  We restricted our
		selected sample
		to images having both entropy values greater than 5, which eliminates some low
		contrast cutouts (see Fig.~\ref{fig:ngc1073} for an example).
		This reduced the number of cutout images to 7,471.

	\item We visually examined the remaining cutout images to eliminate images with obvious
		artifacts such as satellite trails, seams at detector boundaries and heavy CCD charge bleeding.
		Such artifacts were considered likely to distract our reviewers, who are not trained astronomers.
		That eliminated a small number of images, leaving 7,050 cutouts.
		Note we did not try to eliminate all images that included
		stars with diffraction spikes, since that would have discarded
		many NGC star clusters.
		This sample includes 666 different NGC objects.

	\item We retained only a single observation of each object, so the selected sample
		does not include multiple observations of the same piece of the sky.
		We kept the observation that had the largest number of cutout images
		remaining after the above cuts.  If multiple observations had the same
		number of cutouts, we (somewhat arbitrarily) kept the cutouts from the
		observation with a filter central wavelength closest to 550~nm.
\end{itemize}

Our final selected image set includes 2,098 cutout images of 666 NGC objects.
Figure~\ref{fig:samples} shows a sample of the images, which include a wide
variety of morphologies and image textures.
See Appendix~\ref{sec:data-products} for information on the data products that are
available for download.

\begin{figure*}
	\centering
    \includegraphics[width=\textwidth]{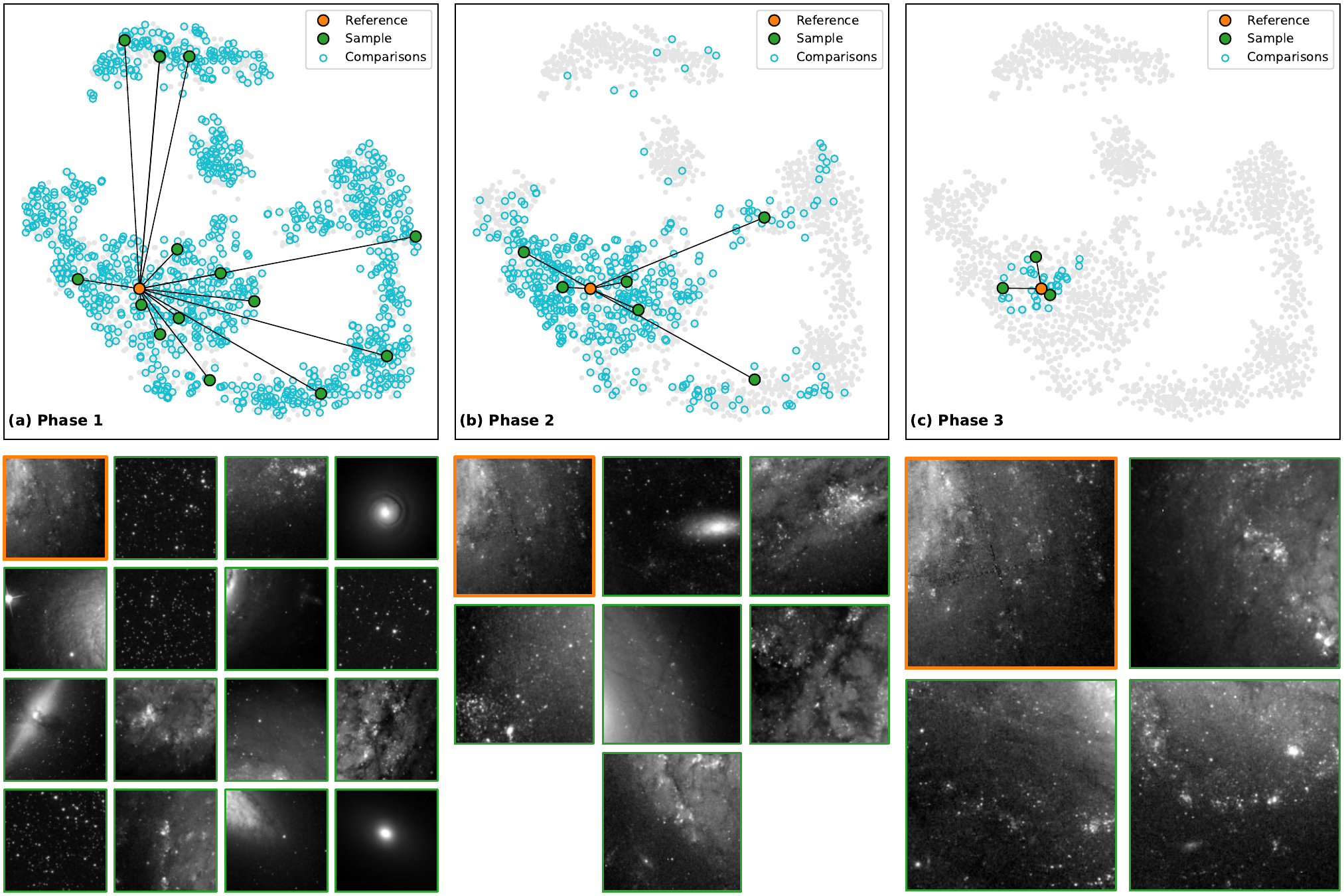}
	\caption{Overview of the three phases of the reviews.  In each column,
	the top plot shows the distribution of of images grouped by similarity
	(section~\ref{sec:tsne}), and the images below show a typical group of
	reference and comparison images.  The plots show all images (gray
	points), the sample reference image (large orange dot), comparison
	images associated with that reference (blue circles), and the
	comparisons from a single randomly selected page (large green dots with
	lines).  (a) In phase 1, all images are compared, and the comparisons
	include a wide variety of images. (b) In phase 2, the focus narrows, and
	the comparison images are more similar to the reference. (c) In phase 3,
	the comparison images are all closely similar to the reference.
\label{fig:3phases}}
\end{figure*}

\begin{deluxetable*}{cccccccc}
\tablecolumns{8}
\tablecaption{HISP Review Phases\label{tab:phases}}
\tablehead{
        \colhead{Phase} &
		\colhead{Pages\tablenotemark{a}} &
		\colhead{\pbox{5em}{\centering Comparisons per~page\tablenotemark{b}}} &
		\colhead{Selection\tablenotemark{c}} &
		\colhead{Repeats\tablenotemark{d}} &
		\colhead{Rotations?\tablenotemark{e}} &
		\colhead{\pbox{5em}{\centering Golden question accuracy\tablenotemark{f}}} &
		\colhead{\pbox{5em}{\centering Mean review time per~page\tablenotemark{g}}}
}
\tablewidth{0pt}
\colnumbers
\startdata
	1 & 151,483 &     15 & All similar  & 1 & no  & 96.3\% &     15.5 s \\
	2 & 123,286 & \phn 6 & Most similar & 3 & yes & 90.8\% & \phn 7.8 s \\
	3 & 107,943 & \phn 3 & Most similar & 3 & yes & 84.7\% & \phn 7.4 s \\
\enddata
\tablenotetext{a}{Number of image groups (pages) presented to reviewers.}
\tablenotetext{b}{Number of comparison images on each page (along with a single reference image).}
\tablenotetext{c}{Reviewer selection task: select all similar images or most similar image.}
\tablenotetext{d}{Number of independent reviewers for each page.}
\tablenotetext{e}{Were images rotated?}
\tablenotetext{f}{Percentage of correct answers for ``golden questions'', where the reference image
is included among comparison images.}
\tablenotetext{g}{Mean time spent by reviewer on each page of images (s).}
\end{deluxetable*}

\section{Image Comparison Procedure} \label{sec:comparisons}

Our reviewers were presented with a reference image and several comparison
images and were asked to identify similar images.  We divided the reviews into
three phases, which differ both in the subsets of images that are compared and
in the decisions that reviewers were asked to make (Fig.~\ref{fig:3phases}).
In phase 1, each reference
image was presented along with 15 comparison images, and reviewers were asked to
identify \emph{all similar images} from among the comparisons.  In phase 2, each
reference image was presented with 6 comparison images, with the reviewer asked
to identify the \emph{most similar image} from the comparisons.  In phase 3,
each reference image is compared with 3 comparison images, again with the
reviewer charged with selecting the \emph{most similar image}.  In each
successive phase, the comparison images are refined to be more similar to the
reference image.

\begin{figure}
    \includegraphics[width=\columnwidth]{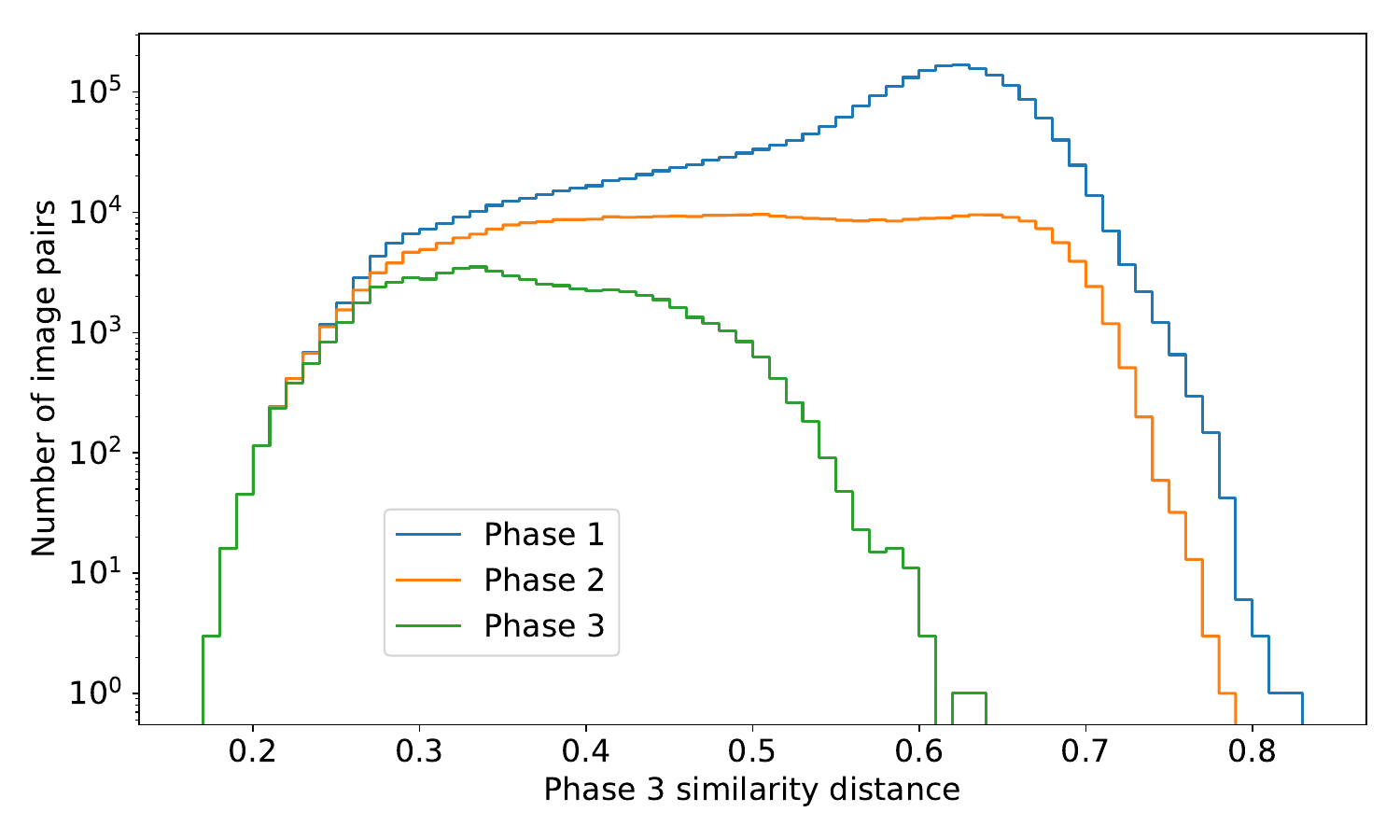}
	\caption{Distribution of similarity distances for pairs of images compared in the three
	phases of the project.  The similarity distance determined using all three phases is used.
	In Phase 1, all image pairs are compared (albeit by only a single reviewer).
	In Phases 2 and 3, more detailed comparisons involving multiple reviewers and requesting
	a more detailed response are focused on successively more restricted pairs of images that
	are chosen to be more likely to be similar.
\label{fig:pair-phase}}
\end{figure}

One way of visualizing the closer focus on similar images in the later phases is
to use the distribution of the similarity distance for the examined pairs.  The
similarity distance is discussed in detail in
Appendix~\ref{sec:similarity-matrix} and is a measure of how similar two images
are (with smaller values meaning more similar).  Figure~\ref{fig:pair-phase}
shows the distribution of this quantity in the three phases.  Note the
logarithmic scale of the $y$-axis; the later phases focus exclusively on more
similar images and exclude very dissimilar images except for rare cases of highly
unusual images having no closely similar counterparts in the sample.

Further details are given in Table~\ref{tab:phases} and in the discussion below.

\begin{figure*}
	\centering
    \includegraphics[width=0.8\textwidth]{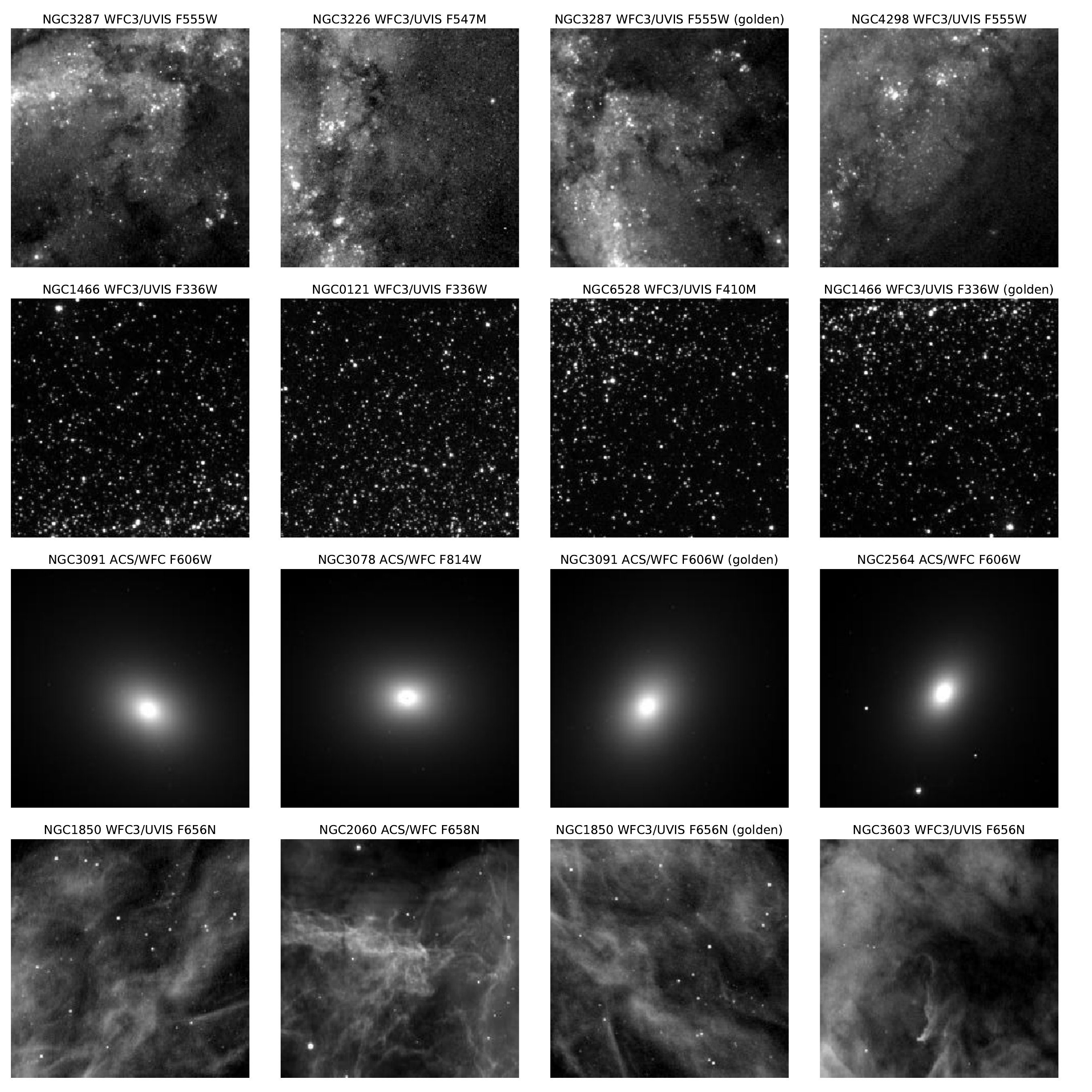}
	\caption{Phase 3 image comparisons for some ``hard'' golden questions.
	Each row has a reference image (left) and 3 comparison images, where one of the
	comparison images is a rotated version of the reference.  
	These were cases where none of the three reviewers correctly selected the
	``golden'' image as the most similar. Certainly the comparison tasks posed
	in phase 3 are very challenging, even for experts.  The matching image is
	indicated in the titles.
\label{fig:hard-golden}}
\end{figure*}

The quality of ``similarity'' being judged by the reviewers is necessarily
somewhat vague.  In our orientation sessions with the reviewers, we discussed
different aspects of similarity (global morphology, morphology of individual
objects, image texture, etc.) and showed some examples with our own choices of
similar images.  We also emphasized that there is no natural orientation for the
images (there is no ``up'') and encouraged them to take that into account when
making choices.  But in the end the choices made by the reviewers reflected
their own independent criteria.  The data from the reviews include identifiers
for the reviewers and also the date of the review, which enables analysis of
reviewer biases and learning.  For the analyses presented in this paper, we
treated all reviewer inputs as equivalent and did not attempt to adjust the
results for reviewer biases.

There are several common characteristics of all three phases. First, 3\% of the
presented comparisons are ``golden questions'', where the reference image itself
is included among the comparison images.  In these cases, the reference image
should always be selected from among the comparison images as similar.  Analysis
of these questions allows us to estimate the accuracy and consistency of
reviews, and can also be used to identify any reviewers who have low accuracy.  Analysis of the performance of our
reviewers on golden questions indicated no serious issues, but were helpful in
providing feedback to individual reviewers in the early learning stages.  Note
that in phases 2 and 3 the images being compared are rotated, which makes
recognizing the image significantly more challenging.  The increasing similarity
of the comparison group also increases the difficulty of identifying golden
questions, particularly in phase 3 (see column~7 in Table~\ref{tab:phases} and
Fig.~\ref{fig:hard-golden}).  See the discussion in section~\ref{sec:biases} for
an example where the golden questions can be useful in the analysis of the data.

Another common feature of the phases is that we avoid comparisons of cutouts
selected from the same NGC object.  Our reasoning for that choice was that we
know adjacent sky regions from the same object are likely to be similar, and we
did not want our final refinement phase to be dominated by comparisons of such
images.  Our ultimate goal is to facilitate the search-by-image functionality,
and archive users searching for similar images would likely be disappointed to
be told that all the most similar sky regions were in the same image as the
original object.  Our approach ensures that we collect sufficient data to answer
the question, ``Which images of other NGC objects are most similar to this
one?''

\subsection{Reviewer Recruitment}

One of the key goals of this project was to support our local community in the time of a global pandemic. To this end, we made contact with a local, well-connected individual who goes by Lou Catelli, or ``The Mayor of Hampden''. He relayed our interest in local workers to various industry professional message boards. We invited all interested participants to an initial orientation session in which we described the project, the system we would be using, and remuneration. We held a few additional orientation sessions for those who could not make the original meeting. We ended up with 30 reviewers who contributed to the project from across the area within walking distance of STScI.

\begin{figure*}
    \includegraphics[width=\textwidth]{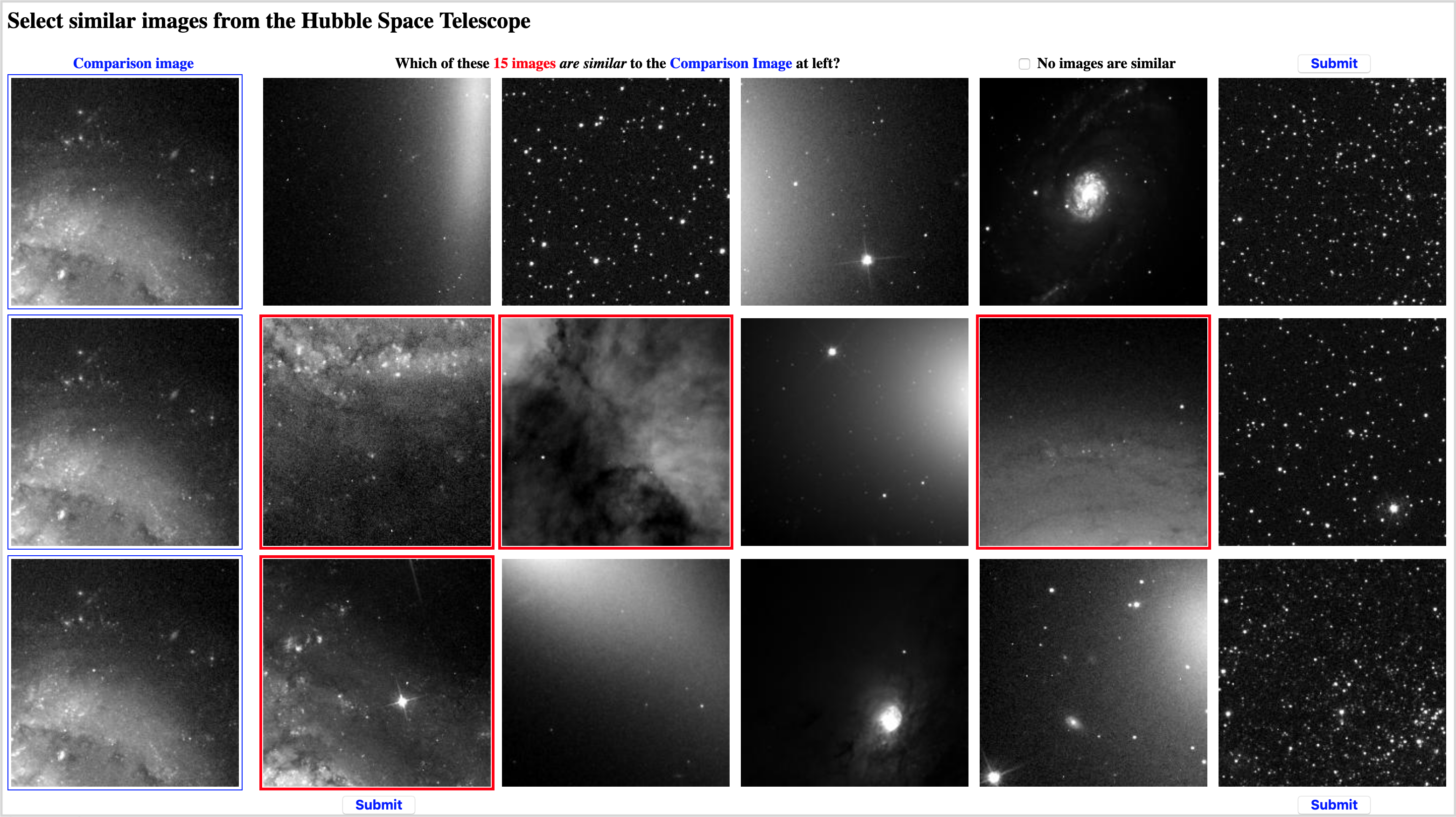}
	\caption{Phase 1 user interface for the reviews.  The reference image is repeated in the
	left column. Users click on one or more of the 15 images to mark them as similar. There is
	also a special checkbox allowing no images to be selected if none are considered similar.
	Selected images are highlighted in red. Clicking a \emph{Submit} button moves on to the
	next page. Users spent an average of 15.5 seconds on each page during this phase.
\label{fig:phase1ui}}
\end{figure*}

\subsection{Phase 1: Broad Comparisons} \label{sec:phase1}

The first phase of comparisons is exhaustive: every image gets compared with
every other image by a single reviewer.  As mentioned above, images from the
same object are not compared. That reduces the number of pairs to be compared by
about 0.35\%, from $N(N+1)/2 = \hbox{2,199,753}$ (where $N=2098$ is the number
of images) to 2,192,069.  The distribution of the actual number of comparisons
for each reference image is peaked close to $N-1 = 2097$ images.

It might appear risky to have only a single reviewer for each pair of images
in phase 1, since that means there is no redundancy to respond to
reviewer errors.  However, there is a very large amount of redundancy
in the matrix of comparisons.  Two images that share similarity
with some subset of images are very likely to be similar to one
another.  Indeed, if a direct comparison is unavailable (or incorrect) for a single
image pair, the set of comparisons to the remaining $\sim2000$ images
for each member of the pair provides a very detailed ``fingerprint'' for those images
and can be used to determine reliably whether the images are similar.
We apply this idea in computing our matrix of similarity distances
(see Appendix~\ref{sec:similarity-matrix}) and discuss it further in the
results (section~\ref{sec:results}).

Comparison pages have a single reference image and 15 comparison images (Fig.~\ref{fig:phase1ui}).  Users
are asked to identify all images from among the comparisons that are considered
to be similar to the reference image.  This means we are using approval voting
\citep{brams1978}, where the reviewer can select as many images as desired (from
zero up to all 15 comparison images).  The images are laid out on the page in
three rows of six images, with the reference image repeated at the beginning of
each row followed by five of the comparison images.  That layout ensures that
the reference image is easily visible even if the reviewer's browser window is
too small to show all the images at once.  After selecting all the similar
comparison images, the reviewer clicks a submit button to record the choices and
proceed to the next page.  To reduce the risk that a reviewer accidentally
clicks the submit button without selecting any images, there is a checkbox on
the page indicating that no images are similar, which must be selected to
proceed when no similar images are identified.  (This is relatively rare, with
$<15$\% of pages having no images selected as similar.)

The comparisons are assumed to be reflexive in this phase: if a review is
performed between reference image $A$ and comparison image $B$, there is no need
to repeat the review with $B$ as the reference and $A$ among the comparisons.
This is likely not to be strictly true, since a cohort of closely similar images
among the comparisons might discourage a reviewer from selecting a slightly less
similar image. But it is expected to be correct for most images, and the quality
and consistency of our results support this expectation.  Subsequent phases
implement the stricter approach of utilizing every image as the reference
(albeit with a smaller subset of comparison images). Those data can be used to
test for and correct any biases introduced by our simplified approach in phase
1.

With this assumption, we can achieve the desired complete set of
cross-comparisons with 147,073 comparison pages.\footnote{It would have required
15 times more comparison pages to complete a full set of comparisons where every
image is used as the reference for cross-comparison with every other image.  The
cost would have been prohibitive considering the limited additional value.} After adding golden questions, our complete phase 1 collection has
151,483 pages of images. For the golden questions, the image that is a duplicate
of the original image is never placed in the first comparison slot on the row
(which would put it immediately beside the reference image and was considered
too easy to recognize.)

\subsection{Phase 2: Closer Comparisons} \label{sec:phase2}

For the second phase of our project, we restrict the comparison subset for each
image to images that appear to be at least moderately similar.  We identify that
subset as follows.  We first compute a ``similarity distance'' matrix $D_{ij}$
between all image pairs $i,j$ in the selected sample from the phase 1 data (see
Appendix~\ref{sec:similarity-matrix} for details).\footnote{For the phase 1
similarity distance analysis, we included the golden questions as part of the
data because approval voting allows all similar images to be marked.  Golden
questions are excluded from phase 2 and 3 analysis since (if answered correctly)
they prevent any other image from being identified as most similar.} Using that
distance matrix, we include the closest $N_n = 200$ neighbors for each image
(approximately 10\% of the full selected sample of 2098 images).  We augment the
neighbors by adding additional images that are closer than a threshold, $D_{ij}
< D_{min}$ where $D_{min} = 0.4$, which adds additional comparison images for
images that are in ``dense'' regions of similarity space with many similar
images close by.  Finally, we expand the comparison group further by including
all image pairs that were selected as similar in the phase 1 reviews.  That
doubtless adds some additional images that are not truly very similar (the
result of errors in the reviews), but it ensures that if there are cases where
our similarity distance model does not accurately capture which images are
similar, those images remain in the comparison cohort for the next phase.

While we aspired to apply the above selection criteria to create the phase 2
collection, in reality there was an error in the construction of the collection
that inadvertently exchanged members of the planned pairs with different cutout
images from the same NGC object.  Approximately half the image pairs were
affected by this issue.  While clearly this mistake was not ideal, the impact on
our results is limited because the phase 2 pairs cast a broad net to include a
large cross section of relatively dissimilar images, and also in many cases the
substituted cutout from an adjacent field in the same object is in fact similar
to the intended image.  The robustness of our algorithm for identifying similar
images via the network of comparisons (see Appendix~\ref{sec:similarity-matrix})
gives us confidence in our results despite the issue.

This is a conservative approach that still leaves a large number of images to be
compared.  On average, every image is compared with 340 other images.
However, reducing the number of pairs by a significant
factor enables us to increase the level of detailed information requested from
the reviewers.  Each comparison page in this phase has a single reference image
with six comparison images.  The reviewer is asked to select the \emph{single
most similar image} from among the comparisons.  That is a more clearly defined
task than identifying all similar images as in phase 1 (where the threshold for
similarity is surely reviewer-dependent).  But it also means that the reviews
are no longer reflexive: if comparison image $B$ is identified as most similar
to reference $A$ among a group, it is not guaranteed that $A$ would be
identified as most similar to $B$ if the roles of those images are exchanged.
That means that the collection of review pages must include both a page where
$A$ is the reference and $B$ is among the comparisons, and another page where
$B$ is the reference and $A$ is among the comparisons.

Another addition in this phase is that we rotate the images being compared
rather than showing all the images in their original orientation.  That is
particularly important for the golden questions where the reference image
appears among the comparisons.  When the golden comparison image is rotated
compared with the reference image, it is significantly more difficult to
recognize.

Finally, we have three independent reviewers identify the most similar image for
each page rather than relying on only a single reviewer as in phase 1.  That
reduces the impact of reviewer errors and inconsistencies.

When all these variations are combined, the phase 2 collection includes 119,695
comparison pages plus 3591 golden questions for a total of 123,286 pages.  The
minimum number of comparison images is 200, but the
typical number is considerably larger with a mean of 340 comparisons.  Each page
is reviewed three times, so the total number of reviews is 369,858.  That is
about 2.5 times larger than the number of reviews in phase 1.  However, each
review requires less time (see column~8 of Table~\ref{tab:phases}), which
allowed us to reduce the payment for each page by a factor of two.  The total
time and cost of phase 2 are about 20\% larger than phase 1.

\subsection{Phase 3: Very Close Comparisons with Rotations} \label{sec:phase3}

For the third phase of our project, we restrict the comparison subset for each
image to images that are closely similar, based on the reviewer input from the
first two phases.  The identification of the subset of images to compare uses an
approach very much like phase 2.  We compute a similarity distance matrix
$D_{ij}$ between all image pairs $i,j$ in the selected sample from a combination
of the phase 1 and phase 2 data (see Appendix~\ref{sec:similarity-matrix} for
details).  Using that distance matrix, we include the closest $N_n = 36$
neighbors for each image (approximately 1.7\% of the full pairwise comparisons
for 2098 images).  We augment those neighbors by adding additional images that
are closer than a threshold, $D_{ij} < D_{min}$ where $D_{min} = 0.21$, which
adds additional comparisons for images that are in ``dense'' regions of
similarity space with many similar images close by.  These parameters were
chosen to collect data on a dense sample of similar images while fitting within
the remaining budget for our project.

On average, every image is compared with 37.5 other images, with only a small
fraction (4.5\%) of the images being compared to more than the minimum set of 36
neighbors. Each comparison page in this phase has a
single reference image with three comparison images.  As in phase 2, the
reviewer is asked to select the \emph{single most similar image} from among
the comparisons.  In order to provide detailed data regarding the influence of
rotation on the perceived similarity, the same subgroup of four images is
presented repeatedly with different relative rotations among the images.  By
showing the groups four times, we acquire similarity comparisons for every pair
of images at all possible relative rotations.

As in the earlier phases, we include golden questions where the reference image
is included among the comparisons.  The golden questions are presented with four
different relative rotations, just like the non-golden questions.  We have three
independent reviewers identify the most similar image for each page, also the
same strategy as in phase 2.

When all these variations are combined, the phase 3 collection includes 26,219
comparison pages plus 787 golden questions for a total of 27,006 pages.  Each
group is repeated with four different rotations, which leads to 108,024 reviews,
but 81 of the groups (0.3\%) have only three rotations owing to various minor
glitches.  That reduces the total number of reviews to 107,943.  Each page is
reviewed by three different reviewers, so the total number of reviews is
323,829.  That is slightly smaller than the number of reviews in phase 2.  The
time required for each review is very comparable to phase 2 (column~8 of
Table~\ref{tab:phases}); the smaller number of comparisons per page compensates
for the significantly greater difficulty of identifying the most similar image
from among a group of images that are often all very similar to the reference
(see Fig.~\ref{fig:hard-golden}).  The total time and cost of phase 3 are about
14\% smaller than phase 2.

\section{Results} \label{sec:results}

It required 87 days to complete all three phases of the reviews, with the time
divided approximately equally among the phases.  Typically there were 2--3 weeks
between the review phases while we verified the data and prepared samples for
the next phase.  Table~\ref{tab:phases} gives a high-level summary of the
contents of the three phases of the project.  It includes the mean time to
review each page of images and the mean accuracy on golden questions.

The golden question accuracy declines in the later phases, where even experts
find it challenging to identify the repeated images (see
Fig.~\ref{fig:hard-golden}).  Nonetheless, the accuracy remains relatively high.
The accuracy for golden questions is lower (as expected) when the comparison
image is rotated compared with the reference image.  In phase 1, the accuracy
for all golden comparisons (none of which are rotated) is $96.3\pm0.3\,\%$.  In
phase 2, the accuracy for unrotated golden comparisons is $93.3\pm0.5\,\%$,
while the accuracy for rotated golden comparisons is $90.0\pm0.3\,\%$.  In phase
3, the accuracy for unrotated golden comparisons is $88.4\pm0.7\,\%$, while the
accuracy for rotated comparisons drops to $83.5\pm0.4\,\%$.

\begin{figure*}[p]
    \includegraphics[width=\textwidth]{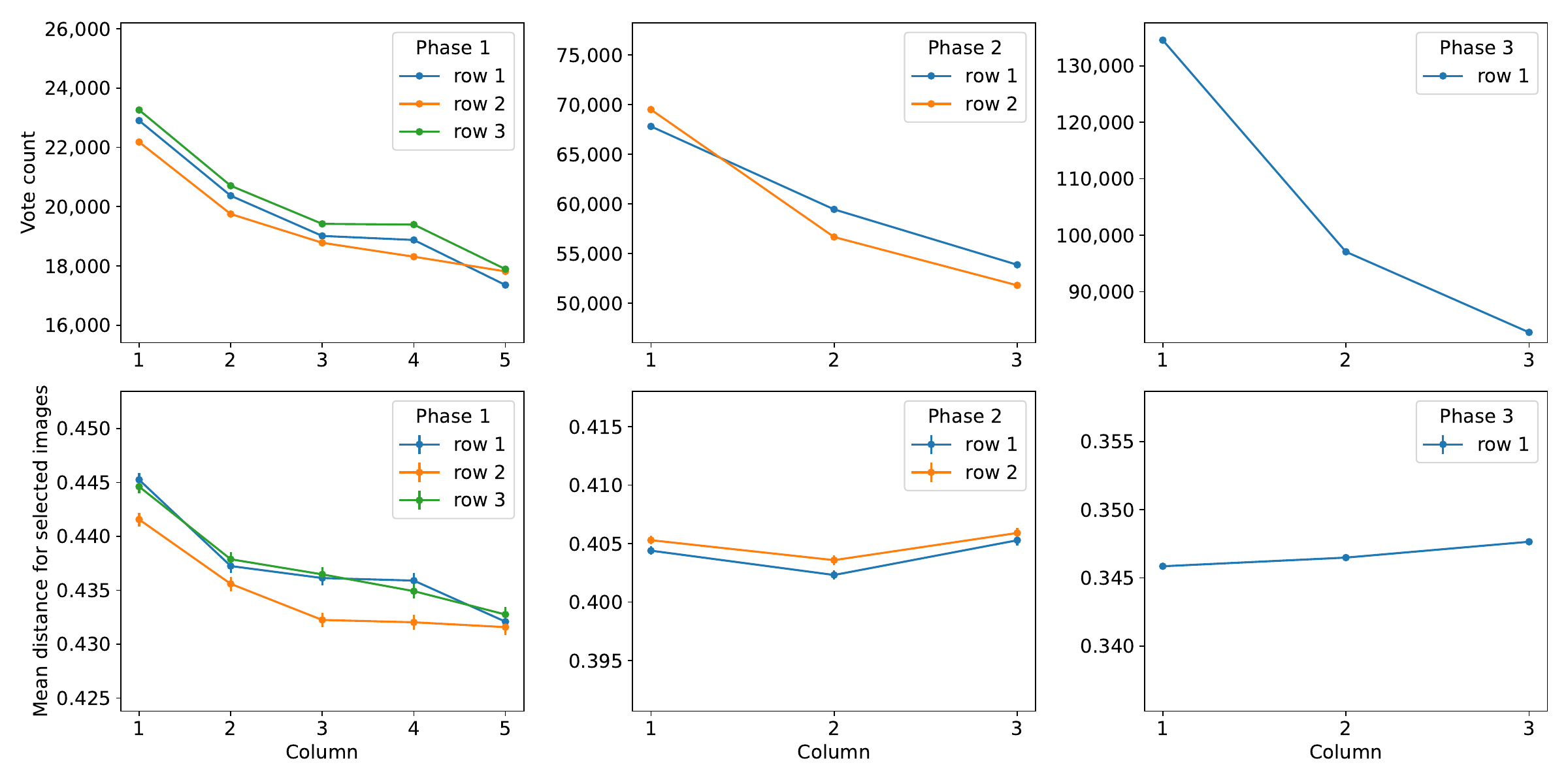}
	\caption{Distribution of reviewer votes as a function of position on page.
	\emph{Top row:} Number of votes for each image as a function of column position for the three phases.  Separate lines
	are shown for each row of images.  There is a significant
	bias in favor of selecting images near the beginning of each row.
	\emph{Bottom row:} Mean similarity distance (using the Phase 3 distances) for the selected images as a function of 
	column position.  The range of variation is much smaller for the computed distances, indicating that the
	distance calculation is much less biased than the reviews.
\label{fig:column-bias}}
\end{figure*}

\begin{figure*}[p]
    \includegraphics[width=\textwidth]{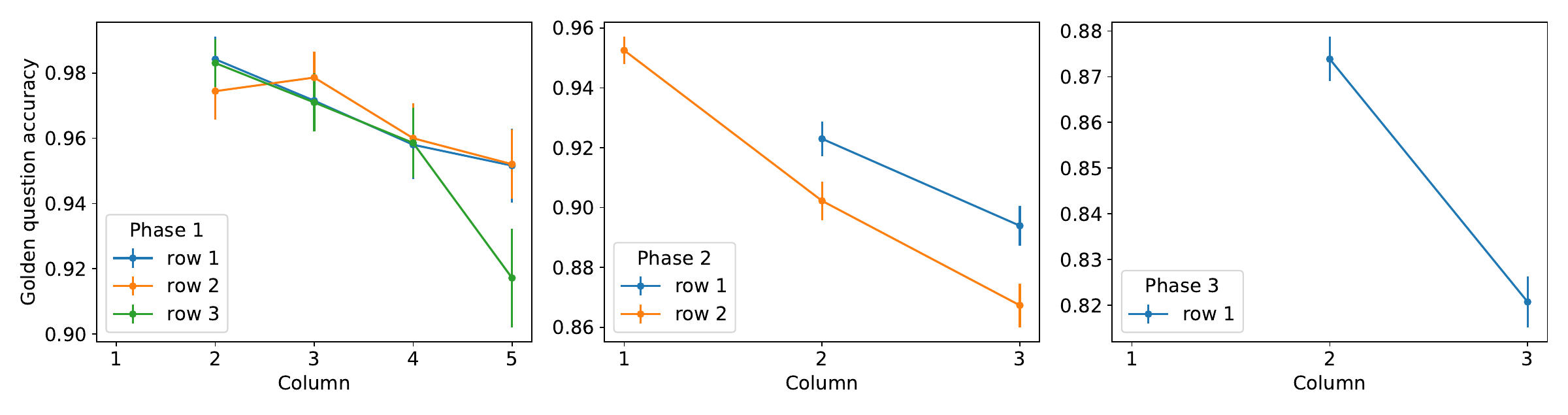}
	\caption{Reviewer accuracy for golden questions as a function of position on page.
	For phases 1 and 3 the golden image (matching the reference image) is
	never placed in column 1; for phase 2 the golden image can appear in
	column 1 of the second row.  The error estimates (from binomial
	statistics) are larger for phase 1 because there are fewer golden
	questions ($\sim350/\mathrm{column}$ in phase 1, compared with
	$\sim1900/\mathrm{column}$ in phase 2 and $\sim4000/\mathrm{column}$ in
	phase 3).
\label{fig:golden-column-bias}}
\end{figure*}

\subsection{Reviewer biases}\label{sec:biases}

The only statistically significant bias that we have discovered in the data is a tendency for
reviewers to select images located closer to the reference image on the web
page.  Figure~\ref{fig:column-bias} shows the distribution of selected images as
a function of the position of the comparison image on the page.  A copy of the
reference image is displayed at the beginning of each row (in column 0).  In all
three phases there is a bias toward selecting images placed in the slots closer
to the reference position.  Since the images are displayed in a random order
(without any knowledge of actual similarity), this can only be the result of
reviewer bias.

Our speculation is that this occurs when a reviewer concludes that no choice
among the comparison images is better than another.  In support of this idea,
the bias increases in the later phases and is most apparent in phase 3, where
sets of very similar comparison images are common (e.g.,
Fig.~\ref{fig:hard-golden}).

The accuracy for golden questions (where the reference image is included as a
comparison) also reflects this bias (Fig.~\ref{fig:golden-column-bias}).  This
is not surprising, particularly in phase 3 where the golden questions can be
challenging (Fig.~\ref{fig:hard-golden}).  This is probably the most useful data
to quantify the bias for phases 2 and 3, where the golden image should (in
principle) always be selected as best.

Fortunately, the effect of this bias on the results is small. The effect is statistically significant, but practically not very important.  
Biases in the selection of similar images --- as well as any
other user errors --- have only a limited effect on our computed similarity
distances (Appendix~\ref{sec:similarity-matrix}).  The similarity distance
between a pair of images relies not on just the review of that single pair, but
on the whole matrix of comparisons with all the other images.  The same images
appear repeatedly on different comparison pages, with a particular image
occurring both as a reference image and as a comparison image in all the
different columns.  A single error has a small impact because the collection of
all other reviews provides a statistical correction for that mistake. That is
apparent in the bottom panels of Figure~\ref{fig:column-bias}, which show the
mean similarity distance for selected pairs as a function of column on the page.
The range of variations in the distance plot is far smaller than the range in
the counts.

Note the mean distance is a little larger in the leftmost bins for phase 1.
That is expected since the bias toward those bins leads to selecting some extra
images having larger distances.  But overall the effect is small. That is partly
due to the robustness of the distance calculation, but also provides support for
the suggestion that these bias errors occur mainly in cases where the images
being selected actually do have similar distances to the other images that could
have been selected.  So whether that pair of images is selected or not in a
particular review, they wind up with the correct distance measure in the end.

\begin{figure}
    \includegraphics[width=\columnwidth]{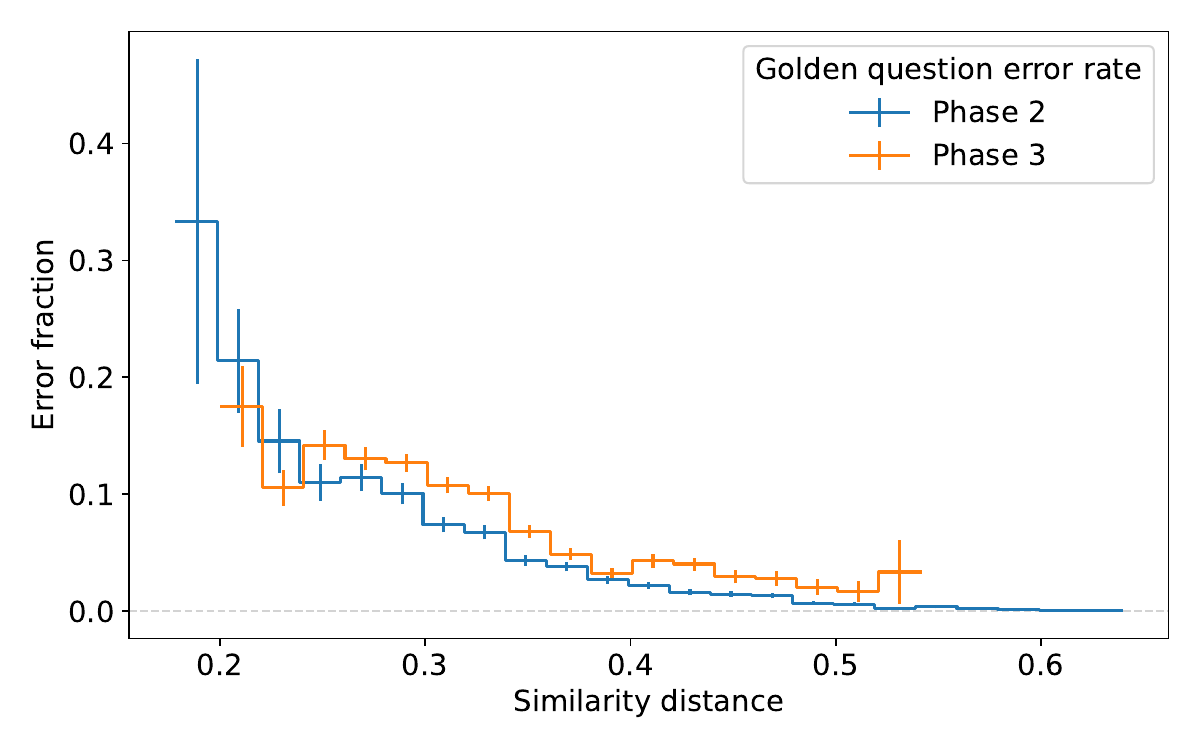}
	\caption{Error rate on golden questions as a function of similarity distance.
	The plot shows the fraction of images as a function of the distance to the
	reference image that are erroneously selected as most similar in phases 2 and 3.
	Most of the images selected incorrectly in the golden question tests have
	distances less than 0.32, whereas the median distance in the sample is 0.60.
\label{fig:golden-dist}}
\end{figure}

The golden questions also provide evidence that errors occur most frequently
when the selected image (if not the golden image) is very similar to the
reference image.  Figure~\ref{fig:golden-dist} shows the fraction of
(non-golden) images erroneously selected in golden questions from phases 2 and 3
as a function of the similarity distance.  Errors are much more common when the
similarity distance is small; errors involving very dissimilar images are
exceedingly rare, regardless of their position on the page.   Fewer than one in
a thousand images with similarity distances greater than the median value of
0.60 are selected in the golden question examples. Only 2\% of image pairs have
distances as close as 0.32, the median for erroneously selected golden question
images.

The bias toward selecting the first image in the row is definitely statistically significant.
If it had been recognized earlier, we likely would have modified the review
design to try to reduce its effects.  However, its influence on the actual
distances that we compute is relatively small.  Mostly it is just a source of
additional measurement noise in our data, as images are positioned randomly on the page.  The network of measurements is
sufficiently robust that it is resistant to this bias (or other reviewer biases
and errors) propagating into the similarity distances.

This bias could probably be reduced or removed with improved data analysis
methods, although any improvement in the final results might not be 
noticeable.  We have not attempted to implement a modified algorithm that takes
into account this bias (or other user biases).

\subsection{Reviewer consistency}\label{sec:consistency}

The consistency between reviewers examining the same collection of images is of interest. We do not
expect perfect agreement since the notion of similarity between two images may depend on many subtle
differences among users (e.g., how an individual reviewer weights large-scale structure versus
small-scale texture).  But at least for some ``easy'' images, the comparison results ought to be
consistent.

For the HISP phase 1 data, there is only a single reviewer for each page of images, so there is not
a simple way to test consistency.  But for the later phases, each page is reviewed by 3 different
reviewers.  For the phase 3 data, each page is presented to reviewers four different times with the
same reference and comparison images but with different rotations of the images. We combine all
reviewer input for the different rotations into a single group that has 12 reviews.

The consistency of reviewer selections for phases 2 and 3 has been assessed using Fleiss' kappa
statistic $K$ \citep{fleiss1971}.  Fleiss' $K$ is a statistical measure of the agreement between
reviewers examining the same collection of images. It requires the same number of ratings for each
item but allows for an arbitrary number of raters as long as they are randomly sampled.  It includes
corrections for random agreements between raters. It also uses the measured distribution of ratings
among categories to correct for category biases. That means it will automatically adjust the
statistic for the first-column bias discussed above.  The Python
\texttt{statsmodels.stats.inter\_rater.fleiss\_kappa} function was used for the calculations.

\begin{figure}
    \includegraphics[width=\columnwidth]{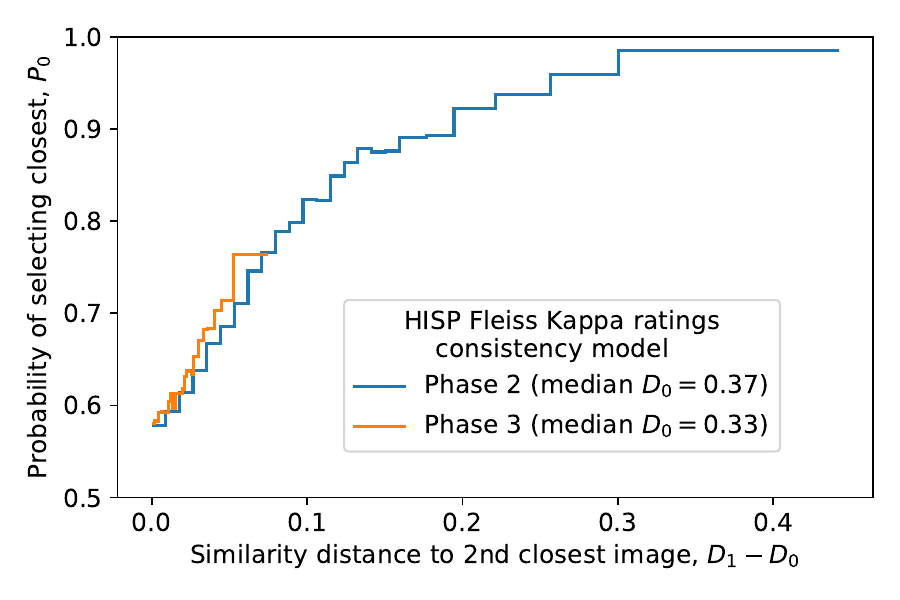}
	\caption{Measure of consistency in ratings from different reviewers derived from Fleiss'
	Kappa. Results are shown for Phase 2 (3 ratings per page) and Phase 3 (12 ratings per page,
	including rotated comparisons).  The $y$-axis is the probability of selecting the best image
	as determined from the final similarity distance matrix.  The $x$-axis is the difference
	between the closest distance ($D_0$) and the second closest ($D_1$).  ``Hard'' examples are
	on the left, and ``easy'' examples are on the right.  The probability of consistent
	selections rises to 98.5\% for the easy cases. Even for hard cases, the probability of
	selecting the closest image is $> 57$\%. (Note that all Phase 3 cases are hard.)
\label{fig:fleiss-kappa}}
\end{figure}

An analysis of all the phase 2 data gives $K = 0.409$.  For the phase 3 data (combining rotations),
$K = 0.204$.  The value of the parameter ranges from 1 (for perfect agreement) to 0 (consistent with
random), with negative values indicating anticorrelations.  There is not a standard score to
evaluate the significance of these $K$ values, but qualitative scales have been suggested that would
make these values range from ``fair'' ($0.21 \le K \le 0.40$) to ``moderate'' ($0.41 \le K \le
0.60$).  These arbitrary labels are considered somewhat dubious because they depend on both the
number of ratings and the number of categories \citep[e.g.,][]{sim2005}.

To get a better idea of the meaning of the $K$ values for our data, we have performed simulations
using a simple model. There is a range of probabilities that the reviewer will choose each object
from a comparison sample. Our single-parameter model assumes that the probability of choosing one of
the $N$ images is proportional to $B^{j}$, with $j=0$ for the most likely choice, $j=1$ for the
second choice, and so on.  If $B=1$ then all categories are equally likely; smaller $B$ indicates a
more strongly peaked distribution.

For the phase 2 data, with 3 raters and 6 choices, $K = 0.409$ corresponds to $B=0.33$.  For the
phase 3 data, with 12 raters and 3 choices, $K = 0.204$ corresponds to $B=0.43$. Averaged over the
entire sample of pages, for phase 2 the best choice is 3.0 times more likely to be selected than the
second choice, and the best choice is selected 67\% of the time.  For phase 3, the best choice is
2.3 times more likely to be selected than the second choice, and the best choice is selected 62\% of
the time.

These averages over the entire phase tell only part of the story on the rating accuracy, however. There are ``hard'' cases 
where it is extremely difficult to select the best choice because multiple images are very similar (Fig.~\ref{fig:hard-golden}),
and there are ``easy'' cases where most of the images are dramatically different.
Figure~\ref{fig:fleiss-kappa} shows the impact of this effect. The samples have been divided into bins
according to how similar the second-closest image compared with the closest image (as determined by our similarity distance, which is
discussed in the next section).  For cases with $x$ well above zero, the second closest image is quite different compared with the closest
image; the result is that the agreement between ratings is excellent, with 98.5\% of the raters choosing the same image. On the other hand,
near $x=0$ the second closest image in the comparison is almost
exactly the same distance as the closest image. The result is is much poorer consistency in ratings.  But even in that case, 58\% of the raters 
pick the same image.

Quantifying consistency of ratings is challenging in this dataset, but we believe there is strong evidence that the results are highly consistent in
cases where the choice is clear, and consistency is good even in the most difficult cases.

\section{Analysis of the similarity matrix} \label{sec:analysis}

The fundamental data product from this project is the collection of reviewer
image similarity comparisons.  But an important derived product, and the true
goal of the project, is a matrix of similarity distances calculated from the
reviews.  Appendix~\ref{sec:similarity-matrix} gives the details of the approach
used to compute a similarity distance matrix $D$ from the reviewer-identified
similarities.

Our similarity distance matrix is designed to capture the ordering of similarity
among groups of images.  If $D_{ij} < D_{ik}$, then images $i$ and $j$ are more
similar than images $i$ and $k$.  We assume that the distance of an image from
itself is $D_{ii}=0$ and that the distance matrix is symmetrical, $D_{ij} =
D_{ji}$.  Otherwise we do not ascribe any meaning to the distance scale.  Our
distances are based on the cosine similarity between patterns of votes for image
pairs, with extreme values of zero and unity.

An essential feature of our approach is that the distances can be computed using
incomplete data.  Every image pair does not have direct reviewer inputs
assessing similarity; this is particularly true in phases 2 and 3, where the
reviewer comparisons are focused on images found to be similar in the earlier
phases. Our similarity distance calculation (eq.~\ref{eqn:distance}) omits
missing data but generates a fully populated output matrix that is (unlike the
original inputs) suitable for use in dimensionality reduction algorithms; see
section~\ref{sec:tsne} below for details.

\begin{figure}
	\centering
    \includegraphics[width=0.9\columnwidth]{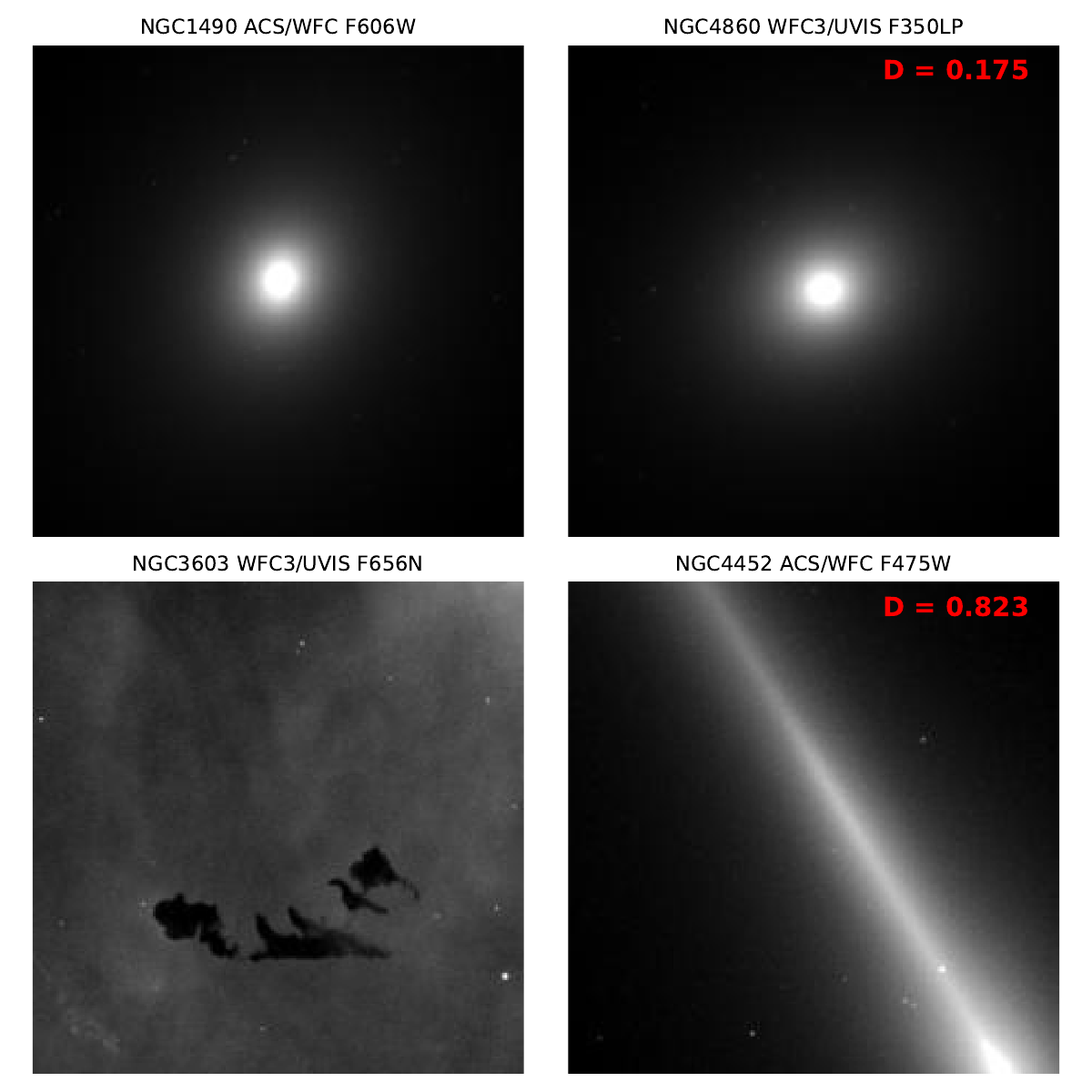}
	\caption{The most similar (top row) and most different (bottom row) image pairs according to
	our similarity distance matrix.
\label{fig:extremes}}
\end{figure}

\begin{figure}
    \includegraphics[width=\columnwidth]{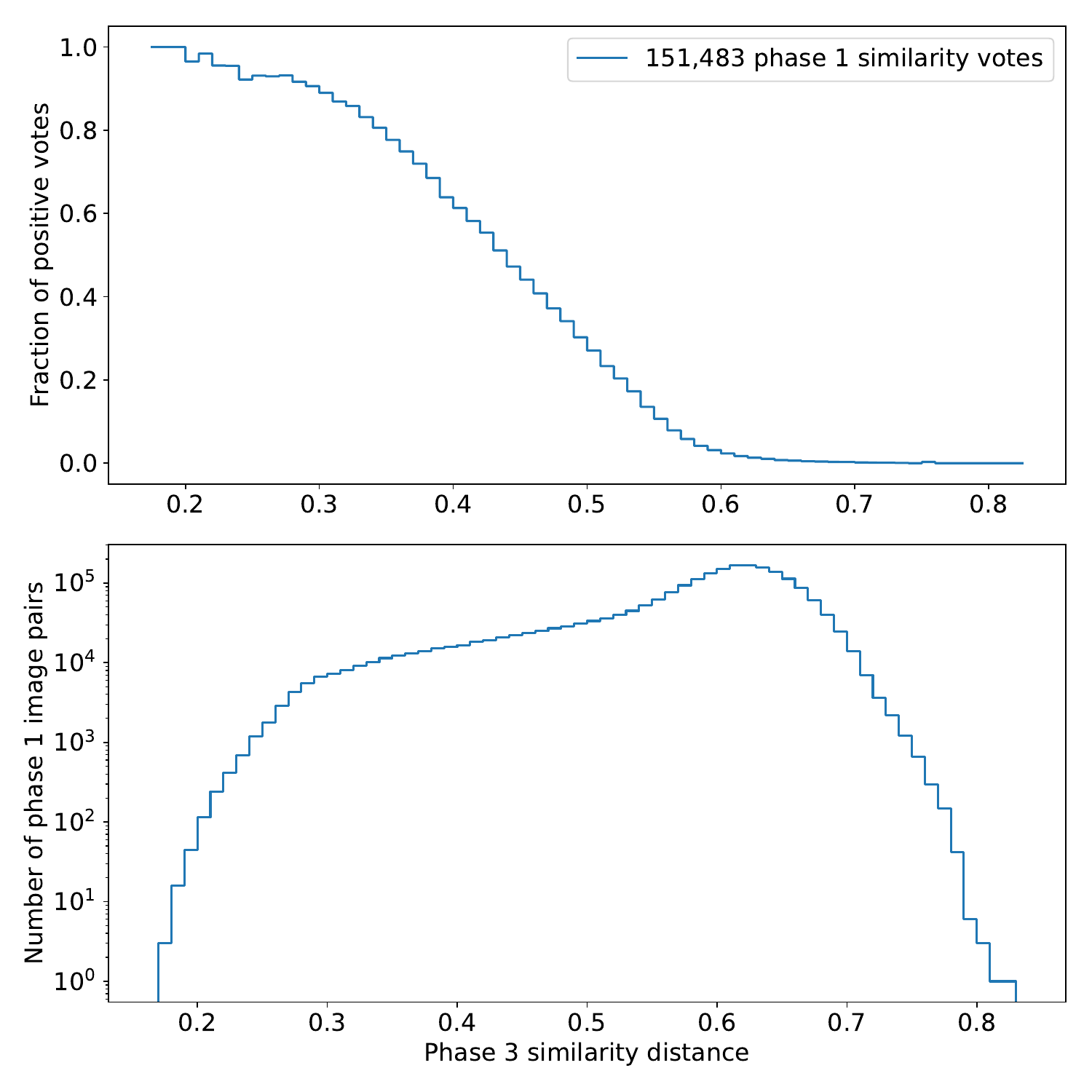}
	\caption{Top panel: Fraction of votes in favor of similarity during phase 1 as a function of the similarity distance.
	Bottom panel: The distribution of similarity distance in the sample (with a logarithmic $y$ scale).
	The similarity distance determined using all three phases is used. Image pairs found to be extremely
	similar (with small similarity distances) are the recipient of a very large fraction of positive
	votes.  Very different images have a very low fraction of positive votes.
\label{fig:phase1-votes}}
\end{figure}

\begin{figure*}
	\centering
    \includegraphics[width=\textwidth]{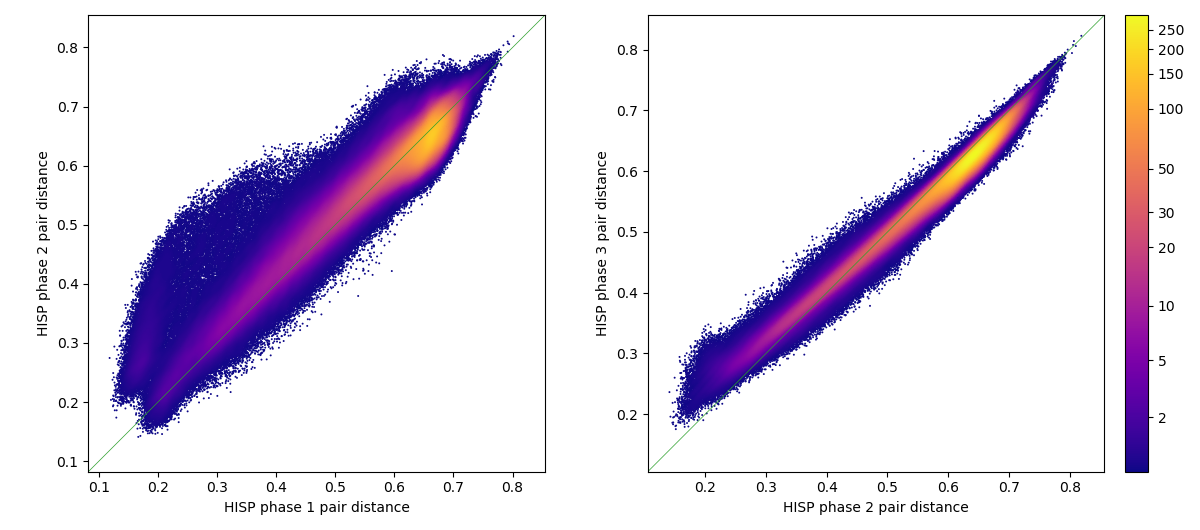}
	\caption{Effect of successive HISP phases on the pairwise similarity distances between
	images.  The left panel compares the distance computed after phase 1 ($x$-axis) with the
	distance after phase 2 ($y$-axis) for the same pair of images.  The right panel compares
	the distance after phase 2 to the distance after phase 3.  The scatter plots are color coded
	by source density to make structure in crowded regions visible (the plots have $\sim2.2\mathrm{M}$ points.)
	The diagonal line indicates equal
	distances, where the additional reviewer data did not change the distances.  The changes between
	phases 1 and 2 are much larger than the changes between phases 2 and 3, indicating a level of
	convergence of the similarity measure.
\label{fig:phase-convergence}}
\end{figure*}

The actual distances range from 0.175 for the most similar pair to 0.823 for the
least similar pair (Fig.~\ref{fig:extremes}).  Figure~\ref{fig:phase1-votes}
shows the distribution of similarity distances for all pairs in the sample, as
well as the votes for all images identified as similar in phase 1.
Figure~\ref{fig:phase-convergence} shows how the computed similarity distances
converge with the phases.  As expected, the changes from phase 2 to phase 3 are
much smaller than the changes from phase 1 to phase 2, indicating convergence in
the later phases.

\begin{figure}[b]
	\vspace{2\baselineskip} 
	\centering
	\includegraphics[width=\columnwidth]{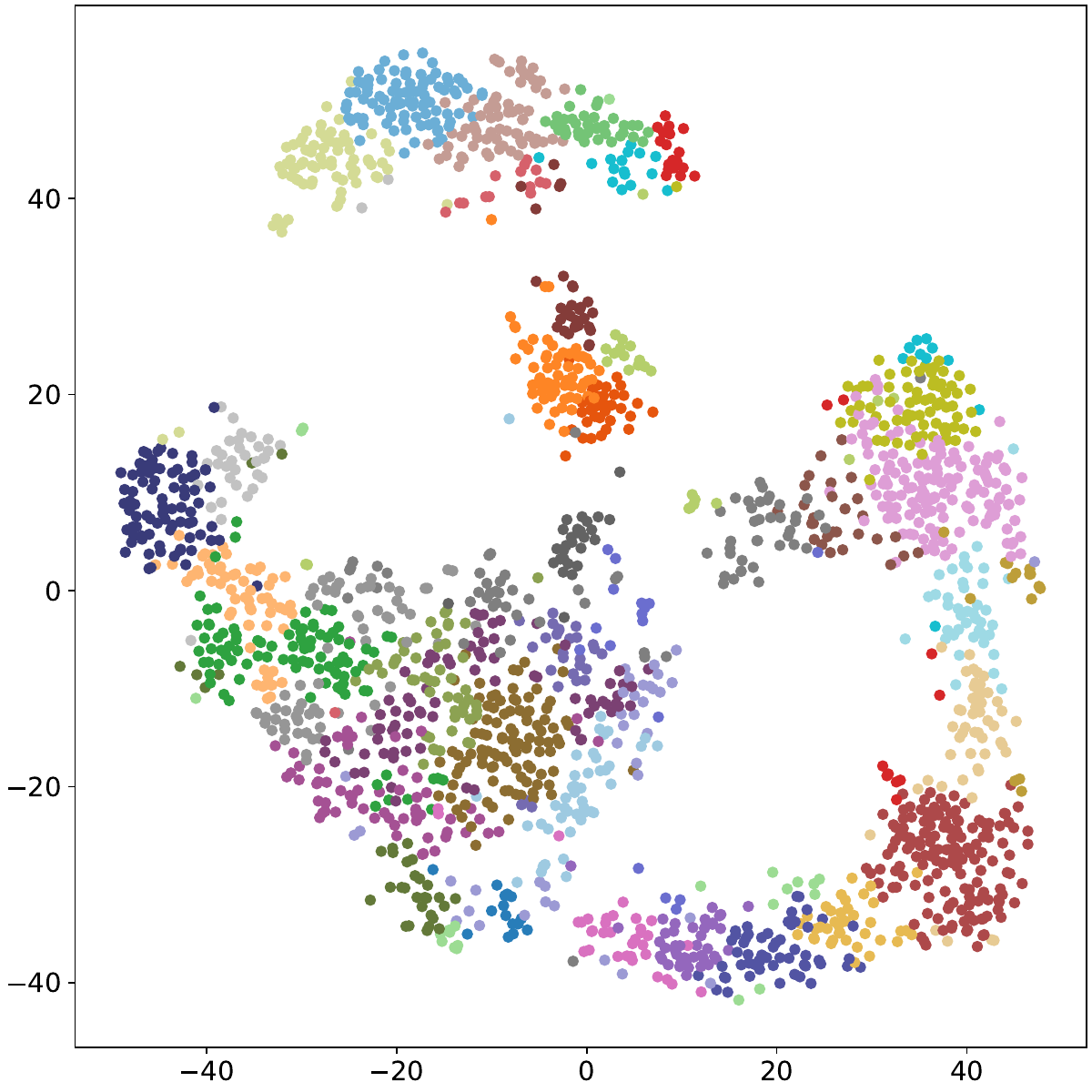}
	\caption{t-SNE mapping of similarity distance for the HISP image sample.  The colors are derived from a
	clustering of the distance matrix using the Ward variance minimization algorithm (implemented in
	\texttt{scipy.cluster.hierarchy.linkage}) and are shown to guide the eye.
\label{fig:tsne}}
\end{figure}

\begin{figure*}
	\centering
	\includegraphics[width=0.9\textwidth]{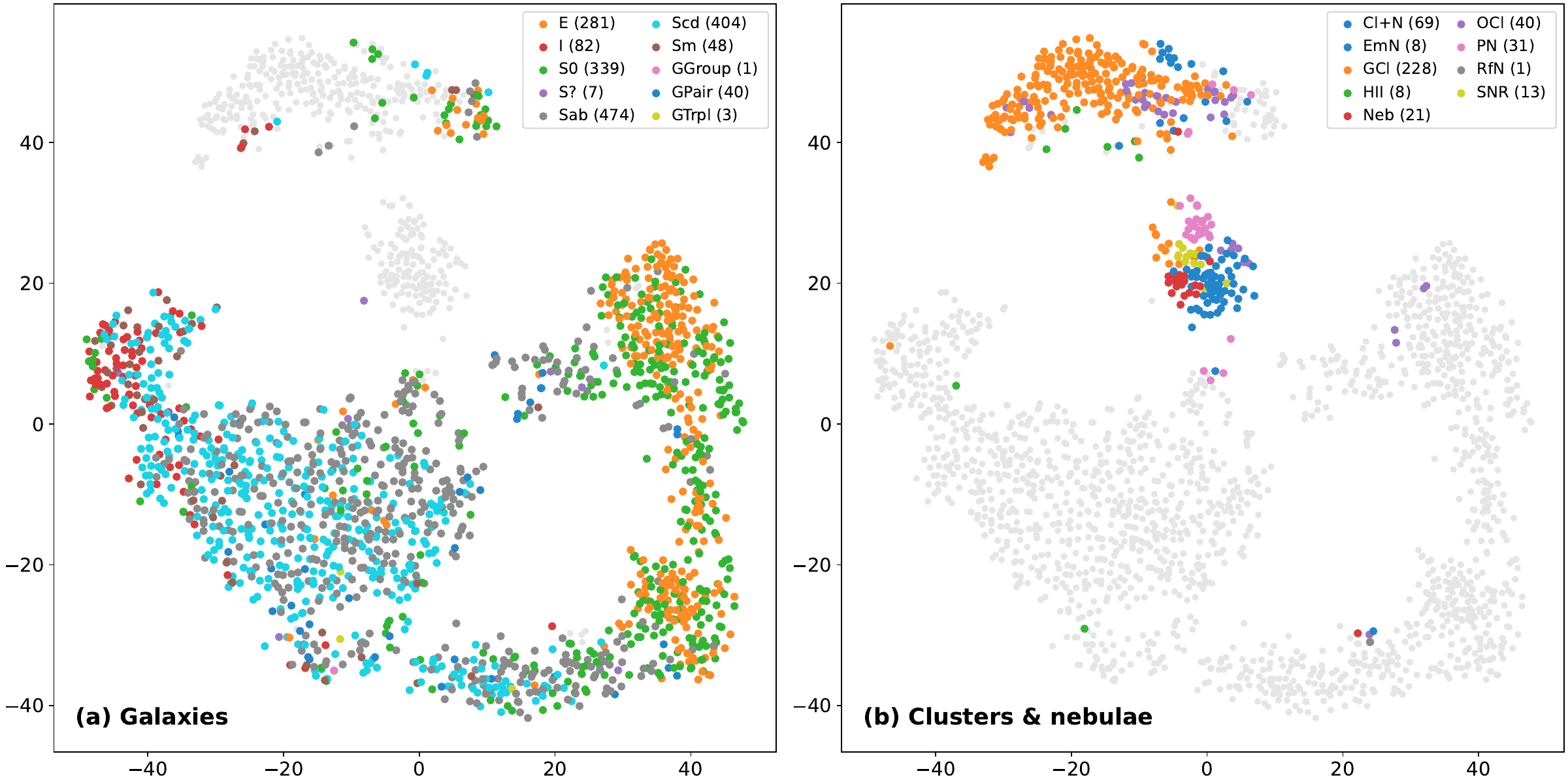}
	\caption{Distribution of NGC object types in t-SNE mapping of similarity distances.
	The left panel shows galaxies, with Hubble types from the NGC catalog condensed into a few
	types.  The right panel shows non-galaxies (star clusters and nebulae).  In both panels, the
	full point distribution is shown by the light gray background points.  There is a distinct
	separation between galaxies and non-galaxies.  There is also a clear separation by object
	type within those categories.
\label{fig:tsne-object-type}}
\end{figure*}

The loneliest image, which is least similar to any other based on the similarity
distance, is in the planetary nebula NGC6720.  Its most similar neighbor, at a
distance of 0.486, is a cutout from the Carina Nebula that has a small dark
dust cloud.  The other images in the sample have an average of 319 neighbors
closer than that distance, and the most crowded region (in a non-descript
region of a nearby spiral galaxy) has 703 neighbors.  Other crowded regions
with lots of similar images include elliptical galaxies (both large and
small).  There is a large dynamic range in the density of images in the
similarity space.

\subsection{Dimensionality reduction using t-SNE}\label{sec:tsne}

We use t-distributed stochastic neighbor embedding \citep[t-SNE;][]{maaten2008}
on the similarity distance matrix to produce a version of the distances that is
suitable for display.  The input is the $2098 \times 2098$ distance matrix,
and the output is a two-dimensional mapping for
each image into a space that preserves the relative distances to nearby
(similar) objects.  We utilize the \texttt{sklearn.manifold.TSNE} Python module
with the \texttt{precomputed} metric, since our input matrix already captures
the distances between points.
The other primary parameters used are \texttt{perplexity=30},
\texttt{n\_iter=5000} and \texttt{learning\_rate='auto'}.

The t-SNE coordinates provide a very useful view of the ``topography'' of our
similarity distances.  They do not preserve the long-range distances between
dissimilar images, but do an excellent job of capturing the short range
clustering.  The t-SNE coordinates themselves are not the primary product
of this work, but they are still a useful visualization tool.  Experiments with
variations in the t-SNE parameters indicate that moderate changes in the chosen
parameters lead to comparable results in the grouping of similar objects.  This
robustness is probably partly attributable to the \texttt{precomputed} metric,
which eliminates the requirement that the algorithm determine the distances from
a (possibly large) collection of features for each image.

The result is shown in Figure~\ref{fig:tsne}. Each point represents a single image, and
nearby points have small similarity distances from one another.  The coloring of
the points is computed by clustering the distance matrix using the Ward variance
minimization algorithm from the \texttt{linkage} and \texttt{fcluster} functions
in the \texttt{scipy.cluster.hierarchy} package.  The \texttt{maxclust}
criterion was used to generate 40 clusters.  It is clear that t-SNE groups these
clusters well, with most clusters well delineated in the t-SNE coordinates.

These similarity coordinates also separate different object types
from the NGC catalog into distinct regions.  Galaxies of all types
(Fig~\ref{fig:tsne-object-type}a) are cleanly separated from nebulae
and star clusters (Fig~\ref{fig:tsne-object-type}b), with almost
no overlap between the objects.  Elliptical galaxies and spirals
are also clearly delineated, with lenticular S0 galaxies forming a
bridge between the two groups.  There are clear separations between
different types of nebulae in the central island of the distribution,
and star clusters lie on their own branch at the top.

This is not surprising given that the morphology of the images is
closely linked to the object types.  But it is still impressive
given that many of the images include only a small portion of the
object.  No information on the object types was used in the selection
of images, the user reviews, or the creation of the similarity
distance matrix.

Visual examination of the images in different regions shows that similar images
are closely grouped, with a gradual change in image morphology and texture as
one moves along the organized structures.  Figure~\ref{fig:tsne-browser1}
displays the impressive results for a sample of 12 images scattered around the
t-SNE diagram.  In each sub-figure, the upper left image is the reference image,
and the images reading across (from left to right and top to bottom) show the
closest four images to the reference using the t-SNE coordinate distances.  The
lower right plot in each panel displays an overview of the t-SNE space, with the
selected images marked with black symbols.

\begin{figure*}[p]
	\gridline{
		\fig{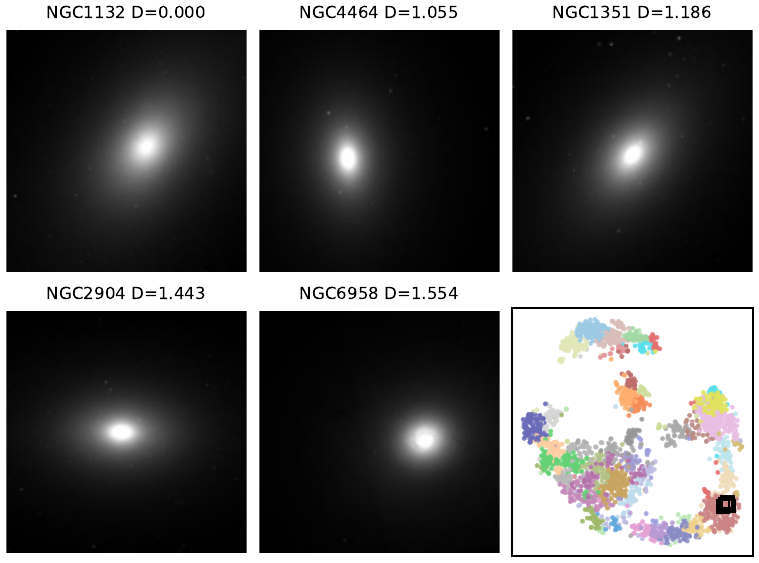}{0.47\textwidth}{(a)}\hskip 0pt plus1filll
		\fig{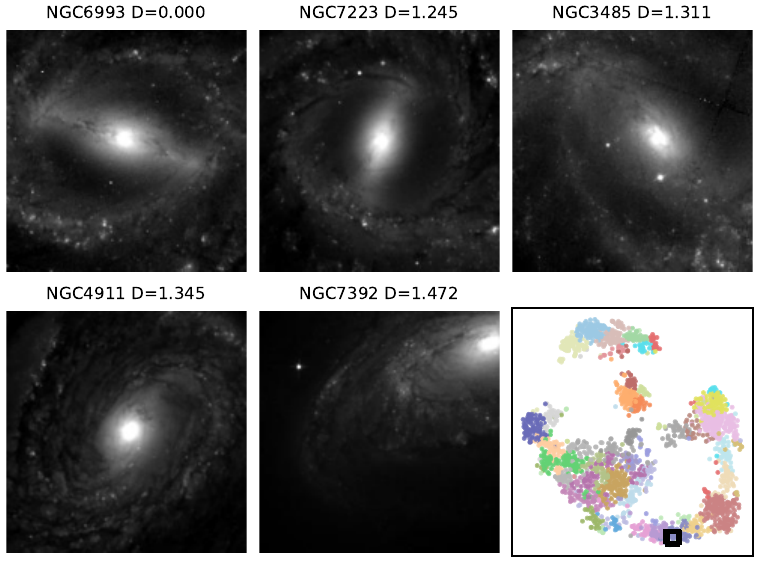}{0.47\textwidth}{(b)}
	}
	\gridline{
		\fig{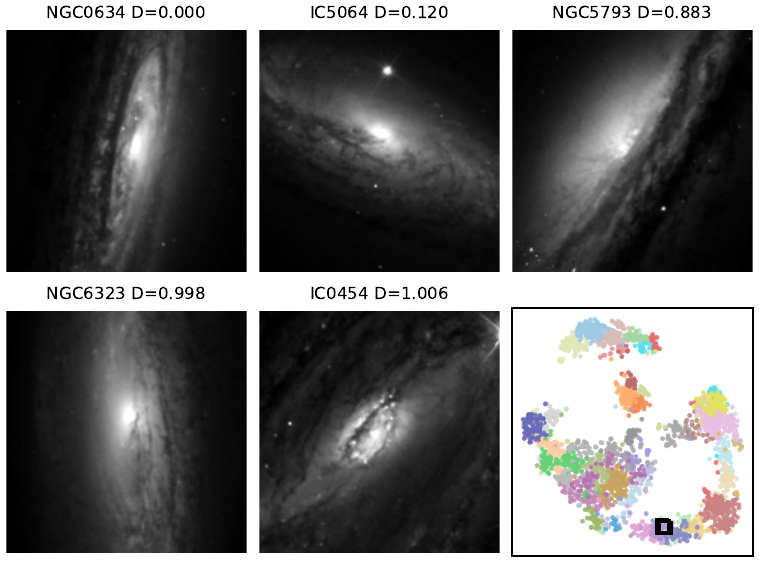}{0.47\textwidth}{(c)}\hskip 0pt plus1filll
		\fig{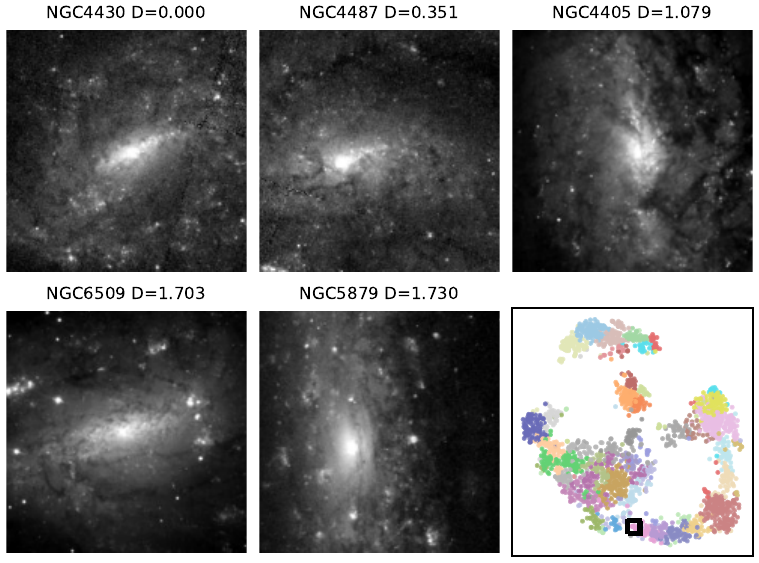}{0.47\textwidth}{(d)}
	}
	\gridline{
		\fig{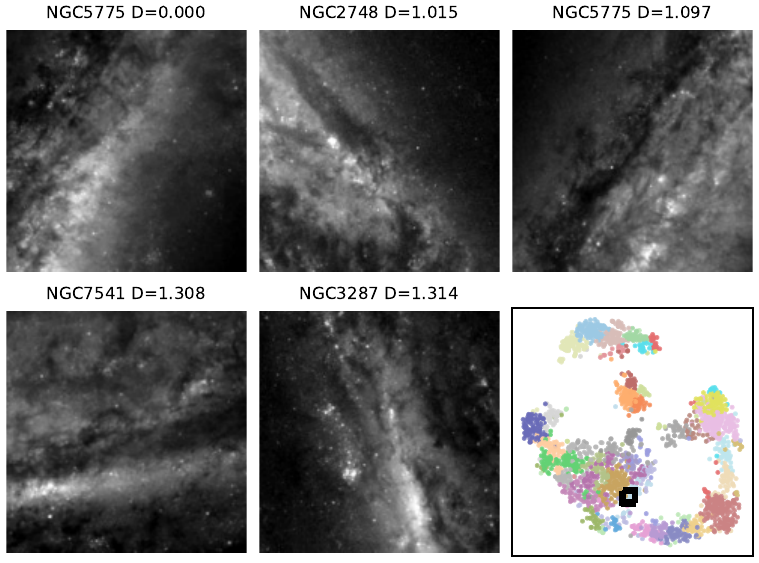}{0.47\textwidth}{(e)}\hskip 0pt plus1filll
		\fig{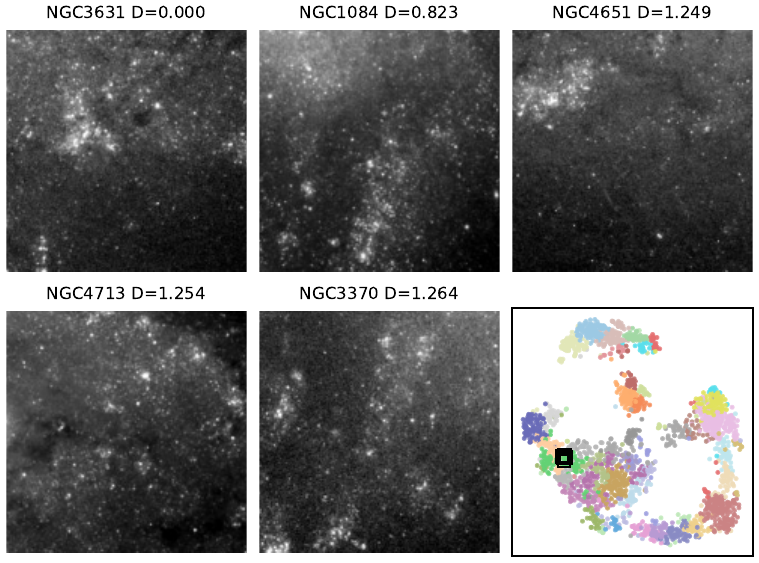}{0.47\textwidth}{(f)}
	}
	\caption{Exploration of the t-SNE parameter space (part 1). Each panel shows a reference image (top left) and the four most similar neighbors. The selected images are marked in black in an overview of the t-SNE space (lower right).
\label{fig:tsne-browser1}}
\end{figure*}

\addtocounter{figure}{-1} 
\begin{figure*}[p]
	\gridline{
		\fig{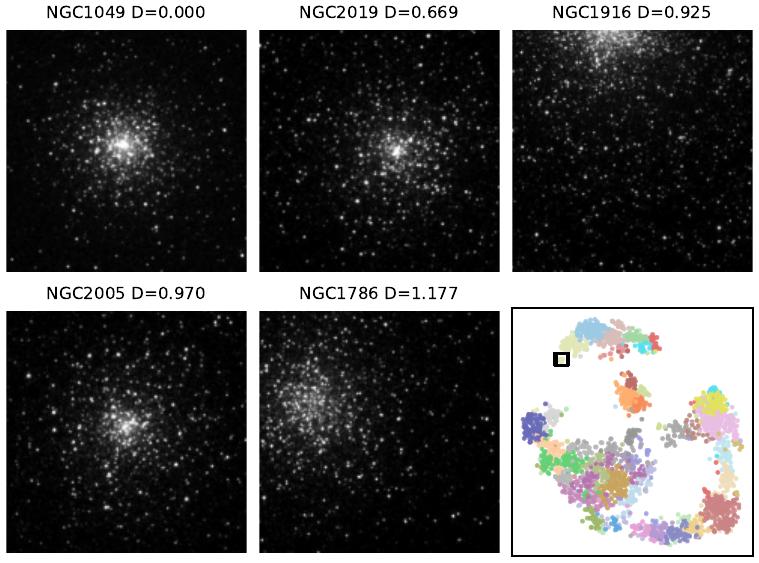}{0.47\textwidth}{(g)}\hskip 0pt plus1filll
		\fig{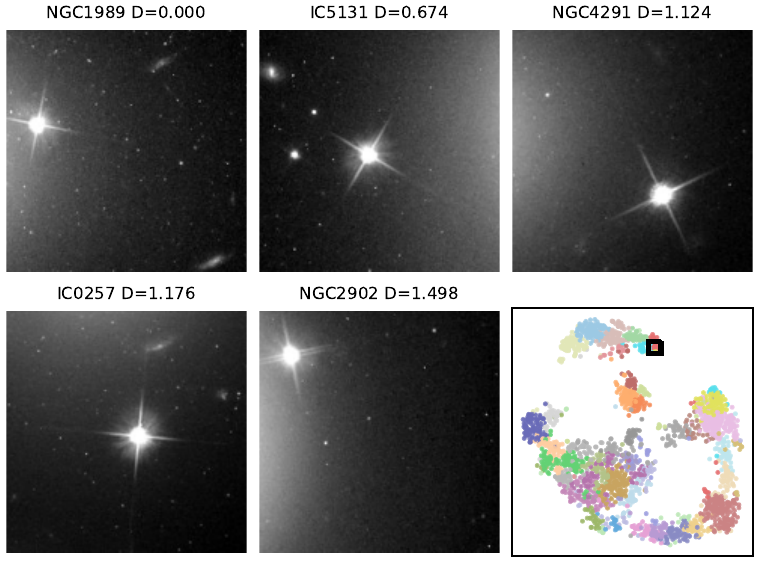}{0.47\textwidth}{(h)}
	}
	\gridline{
		\fig{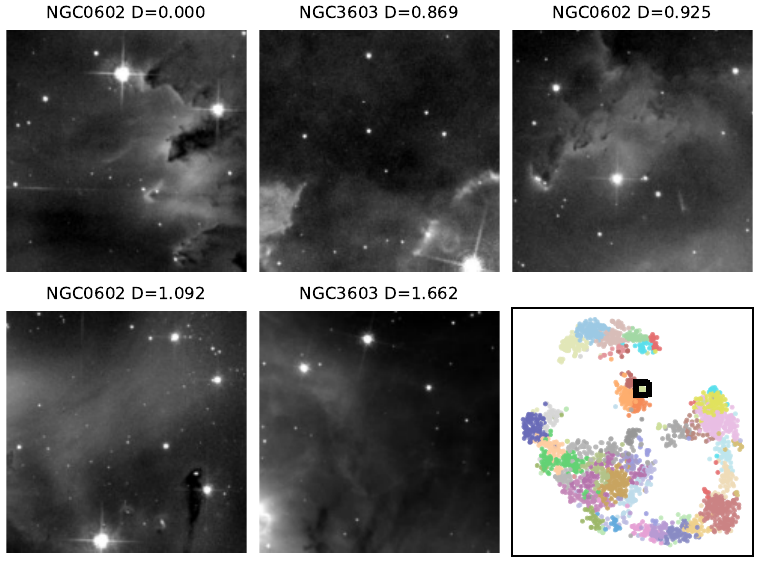}{0.47\textwidth}{(i)}\hskip 0pt plus1filll
		\fig{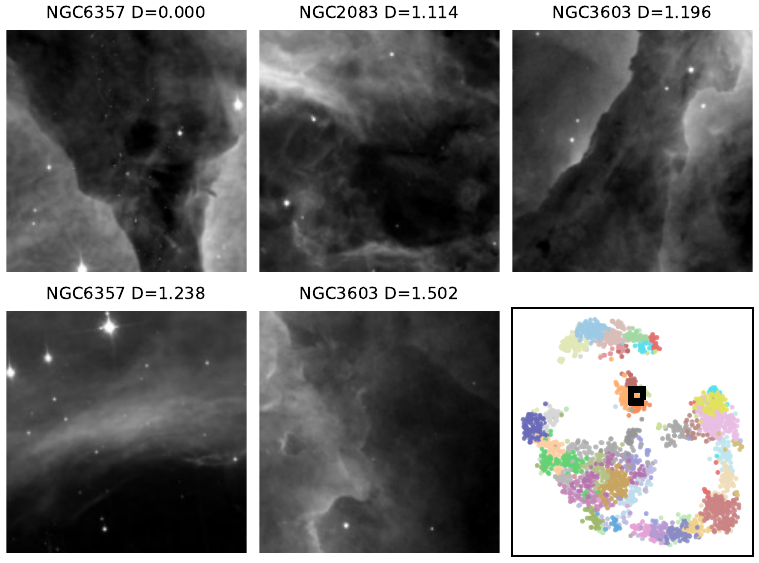}{0.47\textwidth}{(j)}
	}
	\gridline{
		\fig{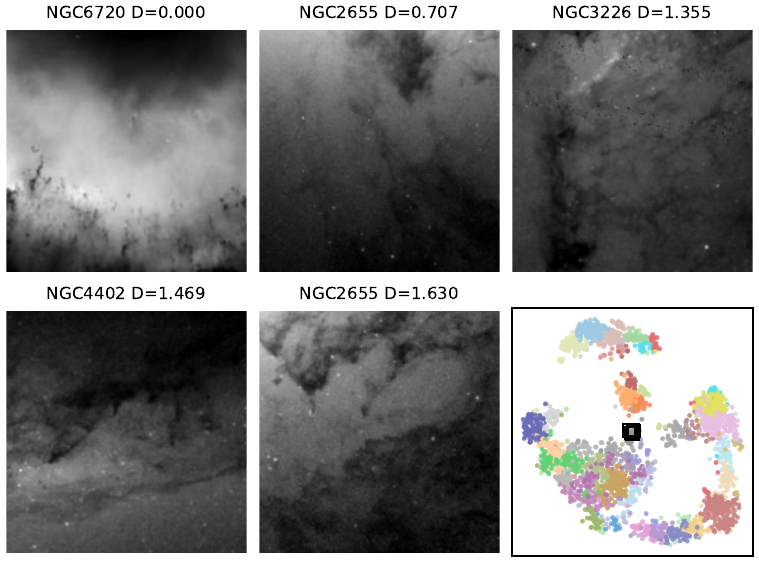}{0.47\textwidth}{(k)}\hskip 0pt plus1filll
		\fig{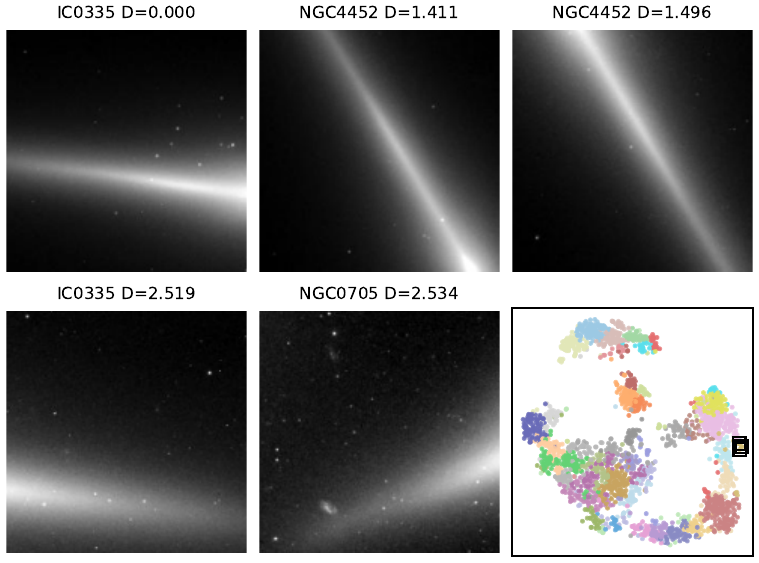}{0.47\textwidth}{(l)}
	}
	\caption{(continued) Exploration of the t-SNE parameter space (part 2).
\label{fig:tsne-browser1b}}
\end{figure*}

The images travel around the outer ring in the t-SNE plot, starting with the
elliptical galaxy branch at the lower right (Fig.~\ref{fig:tsne-browser1}a), and
then moving on through spiral galaxies (Figs.~\ref{fig:tsne-browser1}b--d).  The
large region in the lower left is populated with pieces of large star-forming
spirals, with various amounts of dust (Figs.~\ref{fig:tsne-browser1}e--f).  From
there we jump up to the island of globular clusters
(Fig.~\ref{fig:tsne-browser1}g) and continue on to the upper branch with star
fields.  The density and texture of the star fields changes gradually across the
top, terminating with fields having a few bright stars (with diffraction spikes)
at the upper right (Fig.~\ref{fig:tsne-browser1}h).  The central island in the
middle of the ring is the home of Milky Way star formation regions populated
with clouds of gas and dust (Figs.~\ref{fig:tsne-browser1}i--j) and planetary
nebulae (Fig.~\ref{fig:tsne-browser1}k).  We end our tour at a peninsula on the
far right edge where we find an outcropping of extremely linear, edge-on
galaxies (Fig.~\ref{fig:tsne-browser1}l).

Qualitatively, we find these results to be spectacular. The HISP data capture similarity
between images based on morphology, texture, and other details that are
sometimes difficult even to describe in words (e.g., the lumpy but moderate
contrast star formation regions in Fig.~\ref{fig:tsne-browser1}f and the dusty
bands with sharp edges in Fig.~\ref{fig:tsne-browser1}j).

Our impression, based on a visual examination of the images, is that the t-SNE
distances actually give better results than the similarity matrix distances used
as input to the t-SNE algorithm.  It is difficult to back up this assertion with
data since it relies on extensive visual assessment of the similarity of the
closest images.  We know that the similarity distance matrix captures the
results of the user reviews better (i.e., those distances match the decisions of
our reviewers better than the t-SNE distances).  Our speculation is that the
similarity distance matrix suffers from some overfitting to the data, and that
forcing the results to be projected down to only two dimensions by t-SNE removes
some of the overfitting effects. The t-SNE distances then are a more robust
reflection of the true underlying structure of the similarity space.
Confirmation of this speculation is beyond the scope of this paper.

In summary, we do not claim that our approach for computing the similarity
distances from the reviews is optimal.  It would surely be more useful to have a
meaningful distance scale where the distances between pairs of images predicted
the probability that one image would be preferred over another in a similarity
comparison.  A better algorithm would probably also use cross-validation to
control overfitting.  However, we have found our approach to be simple, understandable, and fast,
and our results as shown in Figure~\ref{fig:tsne-browser1} are fully consistent
with our intuitive interpretation of the meaning of similarity.

\section{Summary and Conclusions} \label{sec:summary}

Our primary goal for this project was to create a large database
of similarity information between astronomical images that can be
used to assess the accuracy of image search algorithms based on
computer vision methods.  We used segments from \emph{HST} images
of NGC objects (galaxies, nebulae, and star clusters) and included
rotated versions of those images to achieve better results for
astronomical images, which lack a natural orientation.  Input was
acquired for 845,170 review samples containing 4 to 16 images over
the three phases of the project.  Our algorithm for extracting a
similarity distance matrix from the human reviews has been shown
to be robust to errors and variance in the input data.  It clearly
produces excellent results (Fig.~\ref{fig:tsne-browser1}).

A secondary goal was to support the local community in Baltimore
during the early months of the Covid-19 pandemic by paying a fair
wage for reviews via the Amazon Mechanical Turk system. We found
the AMT system to be relatively easy to use and manage, and the
results speak to the quality of work that was carried out by our
reviewers.  One benefit of this approach was that we developed a
highly experienced group of citizen scientists who were expert in
both the visual features of the images and the mechanics of doing
reviews.

There is room for possible improvements in the data analysis approach
described here.  The similarity distance matrix could be improved
by having a more easily interpreted scale (e.g., where the different
distances to two objects would predict the likelihood that a user
would select the closest object). It would also be helpful to have
a better approach to limit the impact of overfitting to the data.
Nonetheless, we think our current approach has been shown to be
effective.

We have collected the data for a follow-on project that compares
images of the Martian landscape from the Mars Reconnaissance Orbiter.
That image collection has some differences compared with the current
project.  For example, the Mars images cannot be freely rotated
because the angle of illumination establishes an orientation.  The
results of our Planetary Image Similarity Project will be reported
in a future paper.

All the data products created for this paper are available in
MAST as High Level Science Products:
\dataset[doi:~10.17909/0q3g-by85]{\doi{10.17909/0q3g-by85}}.
The available data (described further in Appendix~\ref{sec:data-products}) include:

\setlist{noitemsep,parsep=0.3\baselineskip,leftmargin=*}
\begin{itemize}
	\item the selected sample of 2,098 HISP images,
	\item the parent sample of 19,916 images from which they were drawn, 
	\item the results of the three-phase citizen science reviews (section~\ref{sec:comparisons}), and
	\item the similarity matrix that we derived, along with the Python code used to compute it
		(Appendix~\ref{sec:similarity-matrix}).
\end{itemize}

\begin{acknowledgments}

This work was supported by the STScI Director's Discretionary
Research Fund (DDRF).  It is based on observations made with the
NASA/ESA Hubble Space Telescope, and obtained from the Hubble Legacy
Archive, which is a collaboration between the Space Telescope Science
Institute (STScI/NASA), the Space Telescope European Coordinating
Facility (ST-ECF/ESA) and the Canadian Astronomy Data Centre
(CADC/NRC/CSA).

Thanks to Sarah Scoles who wrote a wonderful article for \emph{Wired}
magazine\footnote{\url{https://www.wired.com/story/this-citizen-science-gig-pays-people-to-match-space-photos/}}
that described the project and provided important context about the
history of payments for citizen science projects.

Finally, thanks to our citizen scientist reviewers:
Barbara Brenkworth,
Samantha Jade Dill,
James M. Dominguez,
Julia Golonka,
Kimberly Griffith,
Brian Haedrich,
Russell Schofield Huntley,
Katie Jenkins,
Brian Little,
Jake Little,
Jeanine Little,
Laura Mathers,
Lisa McCall,
Kristin Muller,
Brendan O'Connell,
Jonathan Edward Obermaier, Jr.,
Kelly Parsons,
Agatha Pidcock,
Marcy Plimack,
Zakiyyah Seitu,
Geoff Shannon,
Andres Emilio Solano Padilla,
Laurel Stewart,
Patrick Michael Sullivan,
Shaun Michael Vain,
Nicolette Zimmerman,
and Todd Zimmerman.
This project could not have been done without their commitment and
enthusiasm. Special thanks to Lou Catelli (``The Mayor of Hampden'')
for helping assemble our cadre of reviewers from the local community.

\facilities{MAST (HLA)}

\end{acknowledgments}

\clearpage
\appendix

\section{Computing the Similarity Distance Matrix} \label{sec:similarity-matrix}

This appendix describes how we compute a similarity distance matrix from the reviewer-identified
similarities collected by HISP.  We do not claim that this is the ideal approach; certainly there
must be approaches that would make better use of the data, particular in phases 2 and 3 where the
reviewer's task is to select the most similar image.  However, this approach is simple
to implement, practical to compute, and effective in identifying similar images.  We are publishing
the whole dataset so that more sophisticated algorithms can be applied in future work (for example, using models
that incorporate individual reviewer biases and that include estimates of reviewer error rates).

For our approach, we divide the computation of similarity distances into two parts.  First we
compute the \emph{preference matrix} $F$, which gives the probability for every pair of images
that they will be identified as similar. The preference matrix is computed directly from the user
inputs. It is calculated separately for each pair of images, which makes it noisy since it based on
only a few reviews for each image pair. The second step is to compare the preference vectors for
each object to find objects with similar ``fingerprints'' in their $F$ patterns.  That is used to
compute a \emph{similarity distance matrix} $D$ that has much lower noise because it combines
thousands of measurements among many pairs.  The computation of $D$ relies only on $F$ and does not
directly use the user inputs.

We find it useful to keep these two calculations separate rather than combining them into a single
computation, both for computational clarity and also to allow for the possibility of improvement in
the algorithms for either step.

\subsection{The preference matrix}\label{sec:preference-matrix}

In the first phase of HISP, the reviewer is assigned a set of comparison images and asked to identify
all images that are similar to the reference image (see section~\ref{sec:phase1}).  We assume that
there is a symmetrical matrix indicating the probability for each image pair that a reviewer will select
those images as similar.  We also assume that this \emph{preference matrix} $F$ is stable over time and is the
same for different reviewers, and that the preference is independent of which other
images are included in the comparison sample.  (The latter assumptions are surely
oversimplifications.)  We use the binomial distribution plus Bayes theorem to calculate an estimate
of the preference for each pair from the data.

Every review of a page of images results in assessments of similarity between each comparison image
and the reference.  If a comparison image is selected, that is a positive vote for similarity; if a
comparison image is not selected, that is a negative vote against similarity.  We aggregate all the
reviews into two matrices with the positive and negative votes for each pair.
Given $S$ votes in favor of similarity and $N$
votes against similarity for a particular pair, the probability distribution $P(f)$ for the pair
preference $f$ is
\begin{equation}
	P(f|S,N) = \frac{(N+S+1)! }{ N! S!} f^S (1-f)^N \quad .
\end{equation}
That is normalized so that $\int_0^1 df\, P(f|S,N) = 1$.  The mean value of $f$ is
\begin{equation} \label{eqn:pflat}
	\left<f\right> = \int_0^1 df\, f\, P(f|S,n) = \frac{S+1 }{ N+S+2} \quad .
\end{equation}
That is our estimator for the inferred preference value.

This mean value comes from Bayes theorem with a flat prior for $f$, $P(f) = 1$.  But the distribution of $f$
values is not really flat. There are many more dissimilar images than similar images in our sample.  We can
accommodate that by using a power-law prior:
\begin{equation}
	P(f) = (\alpha+1) f^\alpha \quad .
\end{equation}
With this prior the expected mean value of $f$ is $\bar f = (1+\alpha)/(2+\alpha)$.  Inverting
that expression, we get the $\alpha$ value to generate the desired $\bar f$ value:
\begin{equation}
	\alpha = - \frac{1-2 \bar f }{ 1 - \bar f } \quad .
\end{equation}
The nice thing about this prior is that it can be put directly into the $P(f|S,N)$ equation:
\begin{equation}
	P(f|S,N,\alpha) = \frac{\Gamma(N+S+\alpha+2) }{ \Gamma(N+1) \Gamma(S+\alpha+1)} f^{S+\alpha} (1-f)^N \quad ,
\end{equation}
where we recall that $N! = \Gamma(N+1)$ for integer $N$.  Then the mean $\left<f\right>$ becomes
\begin{equation} \label{eqn:prob-prior}
	\left<f\right> = \frac{S+1+\alpha }{ N+S+2+\alpha} \quad . 
\end{equation}
This has the correct limits.  When $S=N=0$ (no data), $\left<f\right> = (1+\alpha)/(2+\alpha) = \bar
f$ as expected. When $S$ and $N$ are large, $\left<f\right> \sim S/(N+S)$, also as expected.  When $\alpha=0$
we recover the result for a flat prior (Eqn.~\ref{eqn:pflat}).

Note that there is no requirement that the prior $\bar f$ be the same for each pair.  In our
implementation we have options for four different priors:
\emph{flat} ($\alpha=0$),
\emph{global} (using the global similarity fraction for all objects),
\emph{object} (using different mean similarity fractions for each object), and
\emph{map} (an option to specify a complete matrix with a separate $\bar f$ value for each pair,
which is used in phases 2 and 3).  For the HISP phase 1 data, the global ${\bar f}$ is 0.134.  We
experimented with these choices and chose to use the \emph{object} prior, which leads to $\bar f$
values ranging from 0.024--0.216 for the 2098 images.  Images with high $\bar f$ values have many
similar neighbors, and those with low $\bar f$ are unusual and have few similar neighbors.  Using a
lower $\bar f$ prior for unusual objects means that we require more evidence to consider comparisons
similar.  This helps compensate for the fact that reviewers do make errors and that those errors are
more likely to impact our analysis for unusual images.

For the phase 1 data, we include golden questions (where the reference image itself is included
among the comparisons) along with the other reviews.  We do not use the self-comparisons, but do
include all the other comparisons.  That is consistent with our simplifying assumption that the
preferences are independent of which other images are included in the comparisons.  It might be
useful to explore the reliability of secondary comparisons in the golden questions to assess the
impact of the other comparison images on reviewer selections, but we have not attempted such an
analysis.

\subsection{The similarity distance matrix}\label{sec:sim-dist-matrix}

The preference matrix described above is independently calculated for each pair of images.  It is
a very noisy estimate of the image properties, with errors in reviewer selections potentially
leading to grossly incorrect values in the matrix.  As a result, this matrix is by itself a low
accuracy, unreliable data product. An enormous number of reviews would be required to reduce the noise for individual elements of
the preference matrix to a useful level.

We can, however, use the preference matrix as input to compute a much more robust and less noisy
product we call the \emph{similarity distance matrix}.  The preference vector for each image
acts as a ``fingerprint'' that identifies that image.  That vector can be compared with vectors for
other images to recognize similarity between even images that have not been directly compared.  For
example, if image $A$ is similar to images $B$, $C$ and $D$, and image $X$ is also similar to those
three images, then image $A$ is likely to be similar to image $X$.  That result becomes stronger
when one considers the \emph{dissimilarity} pattern between images.

We use the cosine distance between preference vectors (the dot product of
the normalized vectors) to quantify the similarity between these vectors.  Because the vectors can have missing
values (image pairs where there is no data), we use a masked cosine distance.  Given the preference
vector values $f(i)$ and $f(j)$ for two objects $i$ and $j$:
\begin{equation} \label{eqn:distance}
	D_{ij} = \frac{ \sum_k W_k f_k(i) f_k(j) }{
		\bigl[ \sum_k W_k f_k^2(i) \, \sum_k W_k f_k^2(j) \bigr]^{1/2} } 
\end{equation}
where the weight $W_k$ is the product of the weights for the two vectors, $W_k = W_k(i) W_k(j)$.
Each of the $W_k(i)$ values is unity for $f_k(i)$ values that are known or zero for values that
should be omitted.

With the above equation we can compute the similarity distance between every pair of images
(including even those that have not been directly compared).  This effectively uses the network of
similar images to get accurate and reliable information on similarity based on noisy and incomplete
data.  The masked cosine distance captures this information simply and is found to be, in practice,
very effective for our dataset.  This simple approach is unlikely to be the optimal method for
determining similarity distances from our reviewer data.  However, we are confident that it is very
useful for this dataset.

One slight modification to the above approach is that we include the diagonal of the preference
matrix (where an object is compared to itself) with unity values, $f_k(k) = 1$.  That is beneficial
to identifying similarities, particularly for unusual images with few neighbors in the similarity
space.  For example, consider the hypothetical case of two images $a$ and $b$ that are identified as similar to one
another but are not similar to any other images in the sample.  By including the diagonal as unity,
the direct comparisons between $a$ and $b$ get included via the term $f_a(b) f_a(a)$. If the diagonal and
its weight $W_k(k)$ are left at zero, that cross term gets omitted and so the one bit of knowledge about
this image pair (that $a$ and $b$ are similar) is ignored.  Since we know that an image by
definition must be maximally similar to itself, it makes sense to assign a preference value of unity.
This change does not have a large effect on our results, but we believe it to be the correct
approach.

\subsection{Phases 2 and 3: Updating the preference matrix}\label{sec:preference-matrix-updates}

For the later phases of HISP, the input data change.  Rather than having reviewers select all similar
images, we ask them to select the most similar image from among the comparisons
(section~\ref{sec:phase2}).  This probably leads to smaller differences among reviewers since (unlike
phase 1) it does not require reviewers to set an arbitrary threshold for what is similar and what is
not.  But it also significantly complicates the data analysis in the later phases.  The reviewer's
choice for a particular pair of images is strongly influenced by which other images are presented
for comparison.  The information that is collected is highly asymmetric: a comparison of images $A$
and $B$ with $A$ as reference is completely different from the same comparison with $B$ as
reference, even if all the other comparison images are the same.  The information being collected is
about \emph{ordering} the similarity distances between the comparisons and the reference.

We do not have a strongly justified algorithm for combining the phase 2 and 3 data with the phase
1 data.  Instead, we have a simple empirical approach that was determined via much experimentation
on the data to perform relatively well.  We take the reviewer inputs in these phases as votes for and
against similarity for the comparison images with the reference.  We do not treat the inputs as
symmetrical: the matrix column of votes for image $A$ only has data where $A$ was the reference
image (not where $A$ was one of the comparisons).  The mean preference for an image pair is
computed with a prior using Equation~(\ref{eqn:prob-prior}).

The major change for the later phases is that the prior in Equation~(\ref{eqn:prob-prior}) is set for
every image pair using the preference matrix determined from the previous phases (using the
\emph{map} option mentioned in section~\ref{sec:preference-matrix}).  In phase 2, the preference matrix
from phase 1 is used to determine the $\alpha$ parameter.  The inputs from phase 2 reviews are then
used to adjust the preferences, with a positive vote increasing the preference and a negative
vote decreasing it.  Similarly, in phase 3 the combined matrix from phases 1 and 2 is used as the
prior.

The computation of the similarity distances from the preference matrix in phases 2 and 3 is exactly
the same as in phase 1 (section~\ref{sec:sim-dist-matrix}).

Our justification for this approach is simply that it appears to work well in practice.  We expect that another
approach might perform as well or better, and we are continuing our search for alternatives.

\section{Data Products}\label{sec:data-products}

All the data products created for this paper are available in
MAST as High Level Science Products:
\dataset[doi:~10.17909/0q3g-by85]{\doi{10.17909/0q3g-by85}}.
The available products 
include the images and associated metadata, details on the comparisons for each phase,  
the results from the three phases of reviews, and some derived products including our similarity matrices.
In the MAST products, all file names have a prefix
\texttt{hlsp\_hisp\_hst\_acs-wfc3-wfpc2\_all\_multi\_v1\_},
which is omitted from the list below.

\begin{itemize}
	\item \texttt{objects.fits}: Properties of the NGC objects used for the sample.
	\item \texttt{observations.fits}: HST archival images (as processed by the HLA) that are sources of the cutout images.
	\item \texttt{cutouts.fits}: Information on all 19,916 JPEG cutout images with identifiers that can be used to get
		additional information on the NGC objects and the HST observations.
		Table~\ref{tab:sample} shows a sample of data from this file.
	\item \texttt{cutouts-selected.zip}: Collection of 2,098 JPEG cutout images selected for the review.
		Note that the comment section in the JPEG files includes the world coordinate system for the image.
	\item \texttt{cutouts-all.zip}: Parent sample of 19,916 JPEG cutout images
		from which the selected images were chosen.  This includes all the selected images as well.
	\item \texttt{phase1-samples.fits}: Table describing the pages of comparison images from phase 1.
	\item \texttt{phase2-samples.fits}: Table describing the pages of comparison images from phase 2.
	\item \texttt{phase3-samples.fits}: Table describing the pages of comparison images from phase 3.
		Table~\ref{tab:phase3_sample} shows a sample of data from this file.
	\item \texttt{phase1-results.fits}: Table with the review results from phase 1.
	\item \texttt{phase2-results.fits}: Table with the review results from phase 2.
	\item \texttt{phase3-results.fits}: Table with the review results from phase 3.
		Table~\ref{tab:phase3_results} shows a sample of data from this file.
	\item \texttt{img-simdist-phs1.fits}: Similarity distance matrix computed from phase 1 results.
	\item \texttt{img-simdist-phs2.fits}: Similarity distance matrix computed from phases 1 and 2 results.
	\item \texttt{img-simdist-phs3.fits}: Similarity distance matrix computed from phases 1, 2 and 3 results.
\end{itemize}

There is a detailed \texttt{readme} file that describes the columns and other properties of the
various files.  Tables~\ref{tab:sample}, \ref{tab:phase3_sample} and \ref{tab:phase3_results} show samples of data from some of the tables.

\clearpage
\movetabledown=1.5in
\begin{rotatetable}
\begin{deluxetable*}{ccccrrlcc}
\tablecaption{HISP Cutout Images\label{tab:sample}}
\tablecolumns{9}
\tablehead{
    \colhead{CutoutID\tablenotemark{a}} & 
    \colhead{ObservationID\tablenotemark{a}} & 
    \colhead{ObjectID\tablenotemark{a}} & 
    \colhead{Name} & 
    \colhead{RA} & 
    \colhead{Dec} & 
    \colhead{Outfile} & 
    \colhead{entropy\tablenotemark{b}} & 
    \colhead{gentropy\tablenotemark{b}} \\
    \colhead{~} &
    \colhead{~} &
    \colhead{~} &
    \colhead{~} &
    \colhead{(J2000)} &
    \colhead{(J2000)} &
    \colhead{~} &
    \colhead{~} &
    \colhead{~}
}
\tablewidth{0pt}
\colnumbers
\tabletypesize{\scriptsize}
\startdata
66 & 22 & 4495 & NGC4496A & 187.9126 & 3.9362 & cutout\_01\_hst\_13695\_40\_wfc3\_uvis\_f606w\_drz.jpg & 5.488 & 7.208 \\
975 & 218 & 1355 & IC4836 & 289.0880 & -60.2004 & cutout\_02\_hst\_14840\_b5\_acs\_wfc\_f606w\_drz.jpg & 6.799 & 6.864 \\
2099 & 444 & 1983 & NGC0602 & 22.3849 & -73.5536 & cutout\_09\_hst\_10248\_05\_acs\_wfc\_f555w\_drz.jpg & 5.683 & 6.620 \\
2632 & 659 & 2305 & NGC1084 & 41.5004 & -7.5856 & cutout\_02\_hst\_11575\_02\_acs\_wfc\_f814w\_drz.jpg & 6.316 & 7.132 \\
3803 & 913 & 2487 & NGC1379 & 54.0053 & -35.4401 & cutout\_01\_hst\_05990\_03\_wfpc2\_f450w\_wf\_drz.jpg & 6.492 & 7.134 \\
4593 & 1065 & 3615 & NGC3226 & 155.8782 & 19.8556 & cutout\_17\_hst\_11661\_01\_wfc3\_uvis\_f547m\_drz.jpg & 5.434 & 6.846 \\
6362 & 1302 & 2830 & NGC1898 & 79.1902 & -69.6563 & cutout\_02\_hst\_13435\_05\_wfc3\_uvis\_f336w\_drz.jpg & 6.148 & 5.000 \\
7342 & 1598 & 3052 & NGC2258 & 101.9187 & 74.4897 & cutout\_01\_hst\_08212\_01\_wfpc2\_f814w\_wf\_drz.jpg & 6.946 & 6.181 \\
7876 & 1771 & 3260 & NGC2661 & 131.4982 & 12.6265 & cutout\_04\_hst\_14840\_3j\_acs\_wfc\_f606w\_drz.jpg & 6.163 & 7.289 \\
8379 & 1957 & 3516 & NGC3059 & 147.4868 & -73.9214 & cutout\_03\_hst\_09042\_f1\_wfpc2\_f606w\_wf\_drz.jpg & 5.601 & 7.029 \\
9385 & 2260 & 3760 & NGC3393 & 162.0977 & -25.1565 & cutout\_04\_hst\_12185\_06\_wfc3\_uvis\_f814w\_drz.jpg & 6.096 & 6.988 \\
\enddata
\tablecomments{Sample of rows from \texttt{cutouts.fits}; full table is available for download.}
\tablenotetext{a}{The \textit{ID} columns identify the cutout, HST observation, and NGC object.  They can be 
used to join to other tables that have detailed information on the object and observation.}
\tablenotetext{b}{The \textit{entropy} and \textit{gentropy} columns give the global and local Shannon entropy for the
cutout image (computed using functions in the Python \texttt{scikit-image} module).  They were used to eliminate cutouts
that have low contrast.}
\end{deluxetable*}
\end{rotatetable}

\clearpage

\begin{deluxetable*}{cccc}
\tablecaption{HISP Phase 3 Samples\label{tab:phase3_sample}}
\tablecolumns{4}
\tablehead{
    \colhead{SampleID} & 
    \colhead{CutoutIDs\tablenotemark{a}} & 
    \colhead{Rotation\tablenotemark{b}} & 
    \colhead{Golden\tablenotemark{c}}
}
\tablewidth{0pt}
\colnumbers
\tabletypesize{\scriptsize}
\startdata
0 & [16633, 7043, 7657, 6875] & [0, 1, 0, 0] & 0 \\
1 & [3956, 4805, 14849, 13723] & [1, 0, 2, 2] & 0 \\
2 & [1496, 18949, 8959, 12624] & [1, 0, 2, 2] & 0 \\
3 & [4264, 13719, 3958, 7663] & [3, 3, 2, 1] & 0 \\
4 & [6858, 17431, 19787, 7268] & [2, 0, 0, 1] & 0 \\
5 & [19307, 19241, 16123, 14849] & [0, 1, 0, 0] & 0 \\
6 & [10795, 2640, 11148, 10133] & [3, 3, 2, 1] & 0 \\
7 & [19303, 13354, 8263, 3209] & [3, 3, 2, 1] & 0 \\
8 & [1917, 7175, 16346, 17998] & [0, 1, 0, 0] & 0 \\
9 & [12278, 8080, 15690, 15450] & [3, 3, 2, 1] & 0 \\
10 & [8125, 19240, 8125, 6706] & [1, 0, 2, 2] & 2 \\
\enddata
\tablecomments{Sample of rows from \texttt{phase3-samples.fits}; full table is available for download.}
\tablenotetext{a}{The \textit{CutoutIDs} column identifies the reference image and three comparison images.}
\tablenotetext{b}{The \textit{Rotation} column indicates the rotation that is applied to the images: 0, 1, 2, and 3
              correspond to counterclockwise rotations by $0^\circ$, $90^\circ$, $180^\circ$ and $270^\circ$ compared
              with the original orientation.}
\tablenotetext{c}{A non-zero value for the \texttt{Golden} column indicates that the reference image is included
              among the comparisons as a test.}
\end{deluxetable*}

\begin{deluxetable*}{cccccc}
\tablecaption{HISP Phase 3 Results\label{tab:phase3_results}}
\tablecolumns{6}
\tablehead{
    \colhead{ResultID} & 
    \colhead{WorkerID\tablenotemark{a}} & 
    \colhead{SampleID\tablenotemark{b}} & 
    \colhead{Work Time\tablenotemark{c}} & 
    \colhead{Submit Time\tablenotemark{d}} & 
    \colhead{Answers\tablenotemark{e}}
}
\tablewidth{0pt}
\colnumbers
\tabletypesize{\scriptsize}
\startdata
89732 & 12 & 0 & 16 & Mon Aug 24 09:17:21 PDT 2020 & -- -- -- \checkmark \\
89730 & 0 & 0 & 3 & Mon Aug 24 10:32:04 PDT 2020 & -- \checkmark -- -- \\
89731 & 8 & 0 & 5 & Mon Aug 24 12:19:30 PDT 2020 & -- -- \checkmark -- \\
115449 & 14 & 5 & 7 & Mon Aug 24 09:31:48 PDT 2020 & -- -- -- \checkmark \\
115450 & 7 & 5 & 34 & Mon Aug 24 09:44:21 PDT 2020 & -- -- -- \checkmark \\
115451 & 0 & 5 & 2 & Mon Aug 24 10:33:51 PDT 2020 & -- \checkmark -- -- \\
32275 & 9 & 10 & 6 & Mon Aug 24 08:31:13 PDT 2020 & -- -- \checkmark -- \\
32276 & 14 & 10 & 5 & Mon Aug 24 09:33:08 PDT 2020 & -- -- \checkmark -- \\
32274 & 8 & 10 & 8 & Mon Aug 24 10:58:49 PDT 2020 & -- -- \checkmark -- \\
\enddata
\tablecomments{Sample of rows from \texttt{phase3-results.fits}; full table is available for download.}
\tablenotetext{a}{The \textit{WorkerID} column uniquely identifies the reviewer.}
\tablenotetext{b}{The \textit{SampleID} column identifies which sample from \texttt{phase3-samples.fits}
              was reviewed. In Phase 3, each sample is reviewed by three different reviewers.}
\tablenotetext{c}{The \textit{Work Time} column indicates the time in seconds required for the review.
              This time can be long if the reviewer took a break.}
\tablenotetext{d}{The \textit{Submit Time} column gives the date and time of the review. This may be helpful
              in exploring drifts in user criteria.}
\tablenotetext{e}{The \textit{Answers} column indicates which comparison image was selected. Only one is
              selected in Phase 3; in Phase 1 multiple images can be selected.}
\end{deluxetable*}

\clearpage

\bibliographystyle{aasjournal}
\bibliography{hisp}

\begin{thebibliography}{}
\expandafter\ifx\csname natexlab\endcsname\relax\def\natexlab#1{#1}\fi
\providecommand{\url}[1]{\href{#1}{#1}}
\providecommand{\dodoi}[1]{doi:~\href{http://doi.org/#1}{\nolinkurl{#1}}}
\providecommand{\doeprint}[1]{\href{http://ascl.net/#1}{\nolinkurl{http://ascl.net/#1}}}
\providecommand{\doarXiv}[1]{\href{https://arxiv.org/abs/#1}{\nolinkurl{https://arxiv.org/abs/#1}}}

\bibitem[{{Astropy Collaboration} {et~al.}(2013){Astropy Collaboration},
  {Robitaille}, {Tollerud}, {Greenfield}, {Droettboom}, {Bray}, {Aldcroft},
  {Davis}, {Ginsburg}, {Price-Whelan}, {Kerzendorf}, {Conley}, {Crighton},
  {Barbary}, {Muna}, {Ferguson}, {Grollier}, {Parikh}, {Nair}, {Unther},
  {Deil}, {Woillez}, {Conseil}, {Kramer}, {Turner}, {Singer}, {Fox}, {Weaver},
  {Zabalza}, {Edwards}, {Azalee Bostroem}, {Burke}, {Casey}, {Crawford},
  {Dencheva}, {Ely}, {Jenness}, {Labrie}, {Lim}, {Pierfederici}, {Pontzen},
  {Ptak}, {Refsdal}, {Servillat}, \& {Streicher}}]{astropy}
{Astropy Collaboration}, {Robitaille}, T.~P., {Tollerud}, E.~J., {et~al.} 2013,
  \aap, 558, A33, \dodoi{10.1051/0004-6361/201322068}

\bibitem[{{Brams} \& {Fishburn}(1978)}]{brams1978}
{Brams}, S.~J., \& {Fishburn}, P.~C. 1978, American Political Science Review,
  72, 831–847, \dodoi{10.2307/1955105}

\bibitem[{{Christiansen} {et~al.}(2018){Christiansen}, {Crossfield},
  {Barentsen}, {Lintott}, {Barclay}, {Simmons}, {Petigura}, {Schlieder},
  {Dressing}, {Vanderburg}, {Allen}, {McMaster}, {Miller}, {Veldthuis},
  {Allen}, {Wolfenbarger}, {Cox}, {Zemiro}, {Howard}, {Livingston}, {Sinukoff},
  {Catron}, {Grey}, {Kusch}, {Terentev}, {Vales}, \&
  {Kristiansen}}]{christiansen2018}
{Christiansen}, J.~L., {Crossfield}, I. J.~M., {Barentsen}, G., {et~al.} 2018,
  \aj, 155, 57, \dodoi{10.3847/1538-3881/aa9be0}

\bibitem[{{Fleiss}(1971)}]{fleiss1971}
{Fleiss}, J.~L. 1971, Psychological Bulletin, 74, 378, \dodoi{10.1037/h0031619}

\bibitem[{{Giavalisco} {et~al.}(1996){Giavalisco}, {Livio}, {Bohlin},
  {Macchetto}, \& {Stecher}}]{hstfaint}
{Giavalisco}, M., {Livio}, M., {Bohlin}, R.~C., {Macchetto}, F.~D., \&
  {Stecher}, T.~P. 1996, \aj, 112, 369, \dodoi{10.1086/118021}

\bibitem[{{Hayat} {et~al.}(2021){Hayat}, {Stein}, {Harrington}, {Luki{\'c}}, \&
  {Mustafa}}]{hayat2021}
{Hayat}, M.~A., {Stein}, G., {Harrington}, P., {Luki{\'c}}, Z., \& {Mustafa},
  M. 2021, \apjl, 911, L33, \dodoi{10.3847/2041-8213/abf2c7}

\bibitem[{{Howard} \& {Ruder}(2018)}]{ssl}
{Howard}, J., \& {Ruder}, S. 2018, arXiv e-prints, arXiv:1801.06146,
  \dodoi{10.48550/arXiv.1801.06146}

\bibitem[{{Koul} {et~al.}(2020){Koul}, {Ganju}, {Kasam}, \& {Parr}}]{astroml}
{Koul}, A., {Ganju}, S., {Kasam}, M., \& {Parr}, J. 2020, arXiv e-prints,
  arXiv:2012.10610, \dodoi{10.48550/arXiv.2012.10610}

\bibitem[{{LeCun} {et~al.}(2015){LeCun}, {Bengio}, \& {Hinton}}]{lecun2015}
{LeCun}, Y., {Bengio}, Y., \& {Hinton}, G. 2015, \nat, 521, 436,
  \dodoi{10.1038/nature14539}

\bibitem[{{Lintott} {et~al.}(2008){Lintott}, {Schawinski}, {Slosar}, {Land},
  {Bamford}, {Thomas}, {Raddick}, {Nichol}, {Szalay}, {Andreescu}, {Murray}, \&
  {Vandenberg}}]{lintott2008}
{Lintott}, C.~J., {Schawinski}, K., {Slosar}, A., {et~al.} 2008, \mnras, 389,
  1179, \dodoi{10.1111/j.1365-2966.2008.13689.x}

\bibitem[{{Lupton} {et~al.}(2004){Lupton}, {Blanton}, {Fekete}, {Hogg},
  {O'Mullane}, {Szalay}, \& {Wherry}}]{lupton2004}
{Lupton}, R., {Blanton}, M.~R., {Fekete}, G., {et~al.} 2004, \pasp, 116, 133,
  \dodoi{10.1086/382245}

\bibitem[{{Mahabal} {et~al.}(2019){Mahabal}, {Rebbapragada}, {Walters},
  {Masci}, {Blagorodnova}, {van Roestel}, {Ye}, {Biswas}, {Burdge}, {Chang},
  {Duev}, {Golkhou}, {Miller}, {Nordin}, {Ward}, {Adams}, {Bellm}, {Branton},
  {Bue}, {Cannella}, {Connolly}, {Dekany}, {Feindt}, {Hung}, {Fortson},
  {Frederick}, {Fremling}, {Gezari}, {Graham}, {Groom}, {Kasliwal}, {Kulkarni},
  {Kupfer}, {Lin}, {Lintott}, {Lunnan}, {Parejko}, {Prince}, {Riddle},
  {Rusholme}, {Saunders}, {Sedaghat}, {Shupe}, {Singer}, {Soumagnac}, {Szkody},
  {Tachibana}, {Tirumala}, {van Velzen}, \& {Wright}}]{mahabal2019}
{Mahabal}, A., {Rebbapragada}, U., {Walters}, R., {et~al.} 2019, \pasp, 131,
  038002, \dodoi{10.1088/1538-3873/aaf3fa}

\bibitem[{{Peek} \& {Burkhart}(2019)}]{pb19}
{Peek}, J.~E.~G., \& {Burkhart}, B. 2019, \apjl, 882, L12,
  \dodoi{10.3847/2041-8213/ab3a9e}

\bibitem[{{Peek} {et~al.}(2020){Peek}, {Jones}, \& {Hargis}}]{peek2020}
{Peek}, J.~E.~G., {Jones}, C.~K., \& {Hargis}, J. 2020, in Astronomical Society
  of the Pacific Conference Series, Vol. 522, Astronomical Data Analysis
  Software and Systems XXVII, ed. P.~{Ballester}, J.~{Ibsen}, M.~{Solar}, \&
  K.~{Shortridge}, 381

\bibitem[{{Rosse}(1850)}]{Rosse1850}
{Rosse}, T. E.~O. 1850, Philosophical Transactions of the Royal Society of
  London Series I, 140, 499

\bibitem[{{Seo} {et~al.}(2023){Seo}, {Kim}, {Lee}, {Han}, {Kim}, {Rey}, \&
  {Song}}]{seo2023}
{Seo}, E., {Kim}, S., {Lee}, Y., {et~al.} 2023, \pasp, 135, 084101,
  \dodoi{10.1088/1538-3873/ace851}

\bibitem[{{Sim} \& {Wright}(2005)}]{sim2005}
{Sim}, J., \& {Wright}, C.~C. 2005, Physical Therapy, 85, 257,
  \dodoi{10.1093/ptj/85.3.257}

\bibitem[{{Skrutskie} {et~al.}(2006){Skrutskie}, {Cutri}, {Stiening},
  {Weinberg}, {Schneider}, {Carpenter}, {Beichman}, {Capps}, {Chester},
  {Elias}, {Huchra}, {Liebert}, {Lonsdale}, {Monet}, {Price}, {Seitzer},
  {Jarrett}, {Kirkpatrick}, {Gizis}, {Howard}, {Evans}, {Fowler}, {Fullmer},
  {Hurt}, {Light}, {Kopan}, {Marsh}, {McCallon}, {Tam}, {Van Dyk}, \&
  {Wheelock}}]{2mass}
{Skrutskie}, M.~F., {Cutri}, R.~M., {Stiening}, R., {et~al.} 2006, \aj, 131,
  1163, \dodoi{10.1086/498708}

\bibitem[{{van der Maaten} \& {Hinton}(2008)}]{maaten2008}
{van der Maaten}, L., \& {Hinton}, G. 2008, Journal of Machine Learning
  Research, 9, 2579.
\newblock \url{http://www.jmlr.org/papers/v9/vandermaaten08a.html}

\bibitem[{Verga(2017)}]{openngc}
Verga, M. 2017, OpenNGC,  The GAVO DC team,
  \dodoi{10.21938/Y.1EJWUD_MQ6B_EDFOVBBW}

\bibitem[{{Whitmore} {et~al.}(2016){Whitmore}, {Allam}, {Budav{\'a}ri},
  {Casertano}, {Downes}, {Donaldson}, {Fall}, {Lubow}, {Quick}, {Strolger},
  {Wallace}, \& {White}}]{whitmore2016}
{Whitmore}, B.~C., {Allam}, S.~S., {Budav{\'a}ri}, T., {et~al.} 2016, \aj, 151,
  134, \dodoi{10.3847/0004-6256/151/6/134}

\bibitem[{{York} {et~al.}(2000){York}, {Adelman}, {Anderson}, {Anderson},
  {Annis}, {Bahcall}, {Bakken}, {Barkhouser}, {Bastian}, {Berman}, {Boroski},
  {Bracker}, {Briegel}, {Briggs}, {Brinkmann}, {Brunner}, {Burles}, {Carey},
  {Carr}, {Castander}, {Chen}, {Colestock}, {Connolly}, {Crocker}, {Csabai},
  {Czarapata}, {Davis}, {Doi}, {Dombeck}, {Eisenstein}, {Ellman}, {Elms},
  {Evans}, {Fan}, {Federwitz}, {Fiscelli}, {Friedman}, {Frieman}, {Fukugita},
  {Gillespie}, {Gunn}, {Gurbani}, {de Haas}, {Haldeman}, {Harris}, {Hayes},
  {Heckman}, {Hennessy}, {Hindsley}, {Holm}, {Holmgren}, {Huang}, {Hull},
  {Husby}, {Ichikawa}, {Ichikawa}, {Ivezi{\'c}}, {Kent}, {Kim}, {Kinney},
  {Klaene}, {Kleinman}, {Kleinman}, {Knapp}, {Korienek}, {Kron}, {Kunszt},
  {Lamb}, {Lee}, {Leger}, {Limmongkol}, {Lindenmeyer}, {Long}, {Loomis},
  {Loveday}, {Lucinio}, {Lupton}, {MacKinnon}, {Mannery}, {Mantsch}, {Margon},
  {McGehee}, {McKay}, {Meiksin}, {Merelli}, {Monet}, {Munn}, {Narayanan},
  {Nash}, {Neilsen}, {Neswold}, {Newberg}, {Nichol}, {Nicinski}, {Nonino},
  {Okada}, {Okamura}, {Ostriker}, {Owen}, {Pauls}, {Peoples}, {Peterson},
  {Petravick}, {Pier}, {Pope}, {Pordes}, {Prosapio}, {Rechenmacher}, {Quinn},
  {Richards}, {Richmond}, {Rivetta}, {Rockosi}, {Ruthmansdorfer}, {Sandford},
  {Schlegel}, {Schneider}, {Sekiguchi}, {Sergey}, {Shimasaku}, {Siegmund},
  {Smee}, {Smith}, {Snedden}, {Stone}, {Stoughton}, {Strauss}, {Stubbs},
  {SubbaRao}, {Szalay}, {Szapudi}, {Szokoly}, {Thakar}, {Tremonti}, {Tucker},
  {Uomoto}, {Vanden Berk}, {Vogeley}, {Waddell}, {Wang}, {Watanabe},
  {Weinberg}, {Yanny}, {Yasuda}, \& {SDSS Collaboration}}]{sdss}
{York}, D.~G., {Adelman}, J., {Anderson}, John~E., J., {et~al.} 2000, \aj, 120,
  1579, \dodoi{10.1086/301513}

\bibitem[{{Zevin} {et~al.}(2017){Zevin}, {Coughlin}, {Bahaadini}, {Besler},
  {Rohani}, {Allen}, {Cabero}, {Crowston}, {Katsaggelos}, {Larson}, {Lee},
  {Lintott}, {Littenberg}, {Lundgren}, {{\O}sterlund}, {Smith}, {Trouille}, \&
  {Kalogera}}]{zevin2017}
{Zevin}, M., {Coughlin}, S., {Bahaadini}, S., {et~al.} 2017, Classical and
  Quantum Gravity, 34, 064003, \dodoi{10.1088/1361-6382/aa5cea}

\end{thebibliography}

\end{document}